\documentclass[
superscriptaddress,aps,preprintnumbers,amsmath,showpacs,amssymb,prd,nofootinbib,preprint]{revtex4-1}

\usepackage[dvipdfmx]{graphicx}
\usepackage{amsfonts}
\usepackage[mathscr]{eucal}
\usepackage{ascmac}
\usepackage{epstopdf}
\usepackage{dcolumn}
\usepackage{bm}
\usepackage{hyperref}
\usepackage{color}

\begin{document}


\def\a{\alpha}
\def\b{\beta}
\def\c{\varepsilon}
\def\d{\delta}
\def\e{\epsilon}
\def\f{\phi}
\def\g{\gamma}
\def\h{\theta}
\def\k{\kappa}
\def\l{\lambda}
\def\m{\mu}
\def\n{\nu}
\def\p{\psi}
\def\q{\partial}
\def\r{\rho}
\def\s{\sigma}
\def\t{\tau}
\def\u{\upsilon}
\def\v{\varphi}
\def\w{\omega}
\def\x{\xi}
\def\y{\eta}
\def\z{\zeta}
\def\D{\Delta}
\def\G{\Gamma}
\def\H{\Theta}
\def\L{\Lambda}
\def\F{\Phi}
\def\P{\Psi}
\def\S{\Sigma}

\def\o{\over}
\def\beq{\begin{eqnarray}}
\def\eeq{\end{eqnarray}}
\newcommand{\gsim}{ \mathop{}_{\textstyle \sim}^{\textstyle >} }
\newcommand{\lsim}{ \mathop{}_{\textstyle \sim}^{\textstyle <} }
\newcommand{\vev}[1]{ \left\langle {#1} \right\rangle }
\newcommand{\bra}[1]{ \langle {#1} | }
\newcommand{\ket}[1]{ | {#1} \rangle }
\newcommand{\EV}{ \ {\rm eV} }
\newcommand{\KEV}{ \ {\rm keV} }
\newcommand{\MEV}{\  {\rm MeV} }
\newcommand{\GEV}{\  {\rm GeV} }
\newcommand{\TEV}{\  {\rm TeV} }
\newcommand{\1}{\mbox{1}\hspace{-0.25em}\mbox{l}}
\def\diag{\mathop{\rm diag}\nolimits}
\def\Spin{\mathop{\rm Spin}}
\def\SO{\mathop{\rm SO}}
\def\O{\mathop{\rm O}}
\def\SU{\mathop{\rm SU}}
\def\U{\mathop{\rm U}}
\def\Sp{\mathop{\rm Sp}}
\def\SL{\mathop{\rm SL}}
\def\tr{\mathop{\rm tr}}

\def\IJMP{Int.~J.~Mod.~Phys. }
\def\MPL{Mod.~Phys.~Lett. }
\def\NP{Nucl.~Phys. }
\def\PL{Phys.~Lett. }
\def\PR{Phys.~Rev. }
\def\PRL{Phys.~Rev.~Lett. }
\def\PTP{Prog.~Theor.~Phys. }
\def\ZP{Z.~Phys. }

\def\dd{\mathrm{d}}
\def\ff{\mathrm{f}}
\def\BH{{\rm BH}}
\def\inf{{\rm inf}}
\def\ev{{\rm evap}}
\def\eq{{\rm eq}}
\def\SM{{\rm sm}}
\def\Mpl{M_{\rm Pl}}
\def\GeV{ \ {\rm GeV}}
\newcommand{\Red}[1]{\textcolor{red}{#1}}

\def\mDM{m_{\rm DM}}
\def\mphi{m_{\text{I}}}
\def\TeV{\ {\rm TeV}}
\def\MeV{\ {\rm MeV}}
\def\Gphi{\Gamma_\phi}
\def\TR{T_{\rm RH}}
\def\Br{{\rm Br}}
\def\DM{{\rm DM}}
\def\Eth{E_{\rm th}}
\newcommand{\lmk}{\left(}  
\newcommand{\rmk}{\right)}
\newcommand{\lkk}{\left[}  
\newcommand{\rkk}{\right]}
\newcommand{\lhk}{\left \{ }  
\newcommand{\rhk}{\right \} }
\newcommand{\del}{\partial}  
\newcommand{\la}{\left\langle} 
\newcommand{\ra}{\right\rangle}

\newcommand{\qel}{\hat{q}_{\rm el}}
\newcommand{\ksplit}{k_{\text{split}}}
\def\GDM{\Gamma_{\text{DM}}}
\newcommand{\half}{\frac{1}{2}}
\def\Gsplit{\Gamma_{\text{split}}}

\def\mg{m_{3/2}}
\newcommand{\abs}[1]{\left\vert {#1} \right\vert}
\def\Im{{\rm Im}}
\def\bea{\begin{array}}
\def\eea{\end{array}}
\def\Mpl{M_{\text{pl}}}
\def\mN{m_{\text{NLSP}}}
\def\Td{T_{\text{decay}}}


\title{
Affleck-Dine Baryogenesis and Dark Matter Production 
after~High-scale~Inflation
}
\author{Keisuke~Harigaya}
\affiliation{Kavli IPMU (WPI), TODIAS, University of Tokyo, Kashiwa, 277-8583, Japan}
\author{Ayuki~Kamada}
\affiliation{Kavli IPMU (WPI), TODIAS, University of Tokyo, Kashiwa, 277-8583, Japan}
\affiliation{Department of Physics and Astronomy, University of California, Riverside, CA, 92507, USA}
\author{Masahiro~Kawasaki}
\affiliation{ICRR, University of Tokyo, Kashiwa, 277-8582, Japan}
\affiliation{Kavli IPMU (WPI), TODIAS, University of Tokyo, Kashiwa, 277-8583, Japan}
\author{Kyohei~Mukaida}
\affiliation{Department of Physics, Faculty of Science,
University of Tokyo, Bunkyo-ku, 133-0033, Japan}
\author{Masaki~Yamada}
\affiliation{Kavli IPMU (WPI), TODIAS, University of Tokyo, Kashiwa, 277-8583, Japan}
\affiliation{ICRR, University of Tokyo, Kashiwa, 277-8582, Japan}
\begin{abstract}
The discovery of the primordial B-mode polarization by the BICEP2 experiment indicates inflation with a relatively high energy scale. Taking this indication into account, we propose consistent scenarios to account for the observed baryon and dark matter densities in gravity and gauge mediated supersymmetry breaking models. The baryon asymmetry is explained by the Afflck-Dine mechanism, which requires relatively low reheating temperature to avoid a sizable baryonic isocurvature perturbation. The low reheating temperature then requires non-thermal production of dark matter to account for the correct relic density of dark matter. Our scenarios can account for the observations of baryon and dark matter density in gravity and gauge mediation and predict some parameters, including the mass of dark matter.

\end{abstract}

\date{\today}
\pacs{98.80.Cq, 95.35.+d, 11.30.Fs, 12.60.Jv}
\maketitle
\preprint{IPMU 14-0093}
\preprint{ICRR-report-677-2014-3}
\preprint{UT-14-17}

\section{\label{sec1}Introduction}

Inflationary cosmology is now almost confirmed 
by the discovery of the primordial B-mode polarization by the BICEP2 experiment~\cite{Ade:2014xna}. 
Then we confront a mystery of the origins of baryon asymmetry and dark matter (DM), 
which are erased by exponential expansion of the Universe during inflation. 
For a consistency of the inflationary cosmology, 
we have to account for the origin of baryon asymmetry and DM after inflation,%
\footnote{
For production of baryon asymmetry and/or DM during inflation, 
see Refs.~\cite{BasteroGil:2011cx, Bastero-Gil:2014oga}. 
}
taking experimental facts into account, including the result of the BICEP2.

The recent result of the BICEP2 experiment implies 
inflation with a relatively high energy scale: 
\beq
 H_{\text{inf}} &\simeq& 1.2 \times 10^{14} \GeV \lmk \frac{r}{0.2} \rmk^{1/2},  \label{H_inf} \\
 r &=& 0.20^{+0.07}_{-0.05} \quad \lmk 68 \% \text{CL} \rmk,
\eeq
where $H_{\text{inf}}$ is the Hubble parameter during inflation, and $r$ is the tensor-to-scalar ratio. 
To 
explain the origins of baryon asymmetry and DM after such a high-scale inflation, 
we focus on 
models of supersymmetry (SUSY). 
In these models, 
the lightest SUSY particle (LSP) is a good candidate for DM, 
and 
the baryon asymmetry can be explained by 
the Affleck-Dine mechanism~\cite{AD, DRT}, 
in which a baryonic scalar field with a flat potential, called the AD field, 
plays an important role. 
However, 
the Affleck-Dine baryogenesis 
after high-scale inflation results in 
a sizable baryonic isocurvature fluctuation~\cite{Enqvist:1998pf, Enqvist:1999hv, 
Kawasaki:2001in, Kasuya:2008xp},%
\footnote{
Axion cold DM is also restricted by the result of BICEP2 and 
isocurvature constraints~\cite{Higaki:2014ooa, Marsh:2014qoa, Visinelli:2014twa, Dias:2014osa}. 
} 
unless the vacuum expectation value (VEV) of the AD field is 
very large during inflation.
The Affleck-Dine baryogenesis with such a large VEV of the AD field 
often requires too low reheating temperature to produce DM thermally.
It is necessary to 
investigate a scenario for non-thermal production of DM.

Moreover, the Affleck-Dine baryogenesis usually predicts 
formation of
a localized lump
composed of condensation of scalar fields carrying
enormously large baryon charge~\cite{Qsusy, KuSh, EnMc, KK1, KK2, KK3}. 
The lump is referred to as a Q-ball~\cite{Coleman}, which is long-lived due to the conservation of 
baryon charge.
Q-balls 
emit quarks from their surfaces and release their charges
into standard model particles~\cite{evap}. 
At the same time, Q-balls may decay into light SUSY particles
and be another source of DM~\cite{EnMc, Fujii:2001xp, Fujii:2002kr, Fujii:2002aj, 
Roszkowski:2006kw, Seto:2007ym, ShKu, Doddato:2011fz, Kasuya:2011ix, Doddato:2012ja, Kasuya:2012mh, KKY}. 

In this paper, 
we construct consistent scenarios to account for the observed baryon and DM densities 
in the cases with and without Q-ball formation in models of gravity and gauge mediation. 
In gravity mediation, 
non-thermal production of DM in a low reheating temperature 
has been investigated in detail in Ref.~\cite{HKMY} (see also Ref.~\cite{Allahverdi:2002nb}), 
where
DM is produced mainly by two mechanisms: 
direct decay of inflaton into SUSY particles~\cite{Moroi:1994rs, Kawasaki:1995cy, 
Moroi:1999zb, Allahverdi:2002nb, Gelmini:2006pw, Kurata:2012nf}
and inelastic scatterings during reheating process~\cite{Allahverdi:2002nb, HKMY}. 
In addition, if Q-balls are formed after the Affleck-Dine baryogenesis, 
DM is also produced from the decay of Q-balls~\cite{EnMc, Fujii:2001xp, 
Fujii:2002kr, Fujii:2002aj, Roszkowski:2006kw, Seto:2007ym, KKY}. 
In this case, 
since the branching fractions of the Q-ball into quarks and gauginos 
are related with each other by a simple relation, 
we may overcome the baryon-DM coincidence problem~\cite{KKY}. 
In gauge mediation, 
the gravitino is the LSP, which is produced by scatterings 
between gluinos and gluons just after the end of reheating 
and is thus related with the reheating temperature~\cite{Moroi:1993mb}. 
Besides, any flat direction other than the one contains Higgs fields 
results in the formation of Q-balls after the Affleck-Dine baryogenesis. 
Then Q-balls provide another source of gravitinos~\cite{ShKu, Doddato:2011fz, Kasuya:2011ix, 
Doddato:2012ja, Kasuya:2012mh}. 
Although gravitinos can be directly produced 
from the decay of Q-balls,
they are mainly produced from the next-to-lightest SUSY particle (NLSP) into which Q-balls decay. 
We find that these scenarios in gravity and gauge mediation can be consistent with 
the observations of baryon and DM densities as well as 
the result of the BICEP2 experiment, 
and have predicted some parameters, including the mass of DM.

This paper is organized as follows. 
In the next two sections, we calculate the baryon density and baryonic isocurvature perturbation 
resulting from the Affleck-Dine baryogenesis, and 
derive an upper bound on reheating temperature
by consistency with observations. 
Then we briefly explain the Q-ball and its properties. 
In Sec.~\ref{models}, we construct scenarios to account for the observed 
baryon and DM density 
in the context of low reheating temperature. 
We consider the cases with and without Q-ball formation, in gravity and gauge mediation. 
Section~\ref{conclusion} is devoted to the summary and conclusion.

\section{\label{ADBG}Affleck-Dine baryogenesis} 

Typical SUSY models, including the Minimal SUSY Standard Model (MSSM), 
contain many flat directions, whose $F$ and $D$ term potentials
are absent in renormalizable levels~\cite{Gherghetta:1995dv}.
Such flat directions are sometimes referred to as AD fields, named after Affleck and Dine. 
The following combination of an up-type right-handed squark and two down-type right-handed squarks, 
denoted as $\phi$, is an example of a flat direction, called $\bar{u} \bar{d} \bar{d}$ flat direction:
\beq
 \tilde{\bar{u}}_1^R = \frac{1}{\sqrt{3}} \phi, \\
 \tilde{\bar{d}}_1^G = \frac{1}{\sqrt{3}} \phi, \\
 \tilde{\bar{d}}_2^B = \frac{1}{\sqrt{3}} \phi,
\eeq
where subscripts and superscripts represent family indices and color indices, respectively. 
In this paper, we do not restrict ourselves to $\bar{u} \bar{d} \bar{d}$ flat direction, 
but use it as an illustration in some cases.

Let us consider classical dynamics of 
an AD field with $B$ (or $B-L$) charge, such as $\bar{u} \bar{d} \bar{d}$ flat direction. 
During inflation,
we assume that the AD field obtains a large VEV.
Non-renormalizable terms, which break 
CP and $B$ (or $B-L$) symmetry in general, are relevant for the dynamics of the AD field due to the large VEV.
After inflation ends, the AD field starts to oscillate 
and rotate in the complex plane 
around the low energy vacuum, 
whose dynamics is far from thermal equilibrium. 
In this way, Sakharov's conditions for baryogenesis~\cite{Sakharov:1967dj} are satisfied
and $B$ (or $B-L$) asymmetry is generated. 
Since the amplitude of the oscillation decreases due to the Hubble expansion, 
the non-renormalizable terms becomes irrelevant and $B$ (or $B-L$) symmetry is approximately restored. 
If the AD field releases its charge into the standard model particles after the sphaleron 
process~\cite{Kuzmin:1985mm}
freezes out, 
the AD field should have $B$ charge to account for the baryon asymmetric Universe. 
On the other hand, 
if the AD field releases its charge before the sphaleron 
process
freezes out, 
the AD field should have $B-L$ charge 
so that the sphaleron process does not wash out the asymmetry. 
In this section, we investigate the Affleck-Dine baryogenesis, taking into account 
the energy scale of inflation given in Eq.~(\ref{H_inf}).

Since we assume that a flat direction obtains a large VEV
in the early Universe, 
higher-dimensional terms should be taken into consideration. 
We assume an R-parity symmetry to avoid catastrophic proton decay, and 
superpotentials such as $W = \bar{u} \bar{d} \bar{d} \supset \phi^3$ disappears. 
Promoting this symmetry into a discrete R-symmetry 
which controls higher-dimensional terms for the flat direction, 
we assume the superpotential of the AD field as
\beq
 W = \frac{\lambda \phi^n}{n \Mpl^{n-3}},
 \label{W}
\eeq
where $\Mpl$ ($\simeq 2.4 \times 10^{18} \GEV$) is the reduced Planck scale, 
$\lambda$ is a coupling constant, 
and $n \ge 4$ is a certain integer which is determined by the R-charge of the AD field. 
For example, $n = 6, 9,12, ...$ for $\bar{u} \bar{d} \bar{d}$ flat direction, 
depending on the discrete R-symmetry.

The AD field has usual soft terms as well as the $F$-term potential 
from the superpotential of Eq.~(\ref{W}) as 
\beq
 V_{\text{s}} &=& m_\phi (\abs{\phi})^2 | \phi |^2
+ \lmk \frac{- \lambda a_g}{n \Mpl^{n-3}} \mg \phi^n + h.c. \rmk, \label{V_s} \\
 V_F &=& \frac{\abs{\lambda}^2}{\Mpl^{2n-6}} \abs{\phi}^{2(n-1)}, \label{V_F}
\eeq
where
$m_\phi (\abs{\phi})$ is the mass of the AD field at the energy scale of $\abs{\phi}$. 
The term proportional to an ${\cal O}(1)$ parameter $a_g$ is an $A$-term mediated by Planck-suppressed interactions, 
and $m_{3/2}$ is the mass of the gravitino. 
Hereafter, we set the phase of the AD field and the SUSY-breaking F-term 
such that $\Im [ \lambda ] = 0$ and $\Im [ a_g] = 0$. 

The AD field acquires Hubble induced terms during inflation~\cite{DRT}, 
because inflation is associated with non-zero vacuum energy $V = 3 H^2 \Mpl^2$, 
which largely breaks SUSY. 
We write these Hubble-induced terms as 
\beq
V_{\text{H}} &=& c_H H^2 \abs{\phi}^2
+ \lmk \frac{ - \lambda a_H}{n \Mpl^{n-3}} H \phi^n + h.c. \rmk, 
\label{V_H}
\eeq
where $c_H$ and $a_H$ are ${\cal O}(1)$ constants. 
Hubble-induced $A$-terms are absent during inflation
if the field which has a non-zero $F$-term during inflation is charged under some symmetry
and its VEV is less than the Planck scale during inflation~\cite{Kasuya:2008xp}.
These conditions are usually 
satisfied for models of high-scale inflation in supergravity~\cite{Kawasaki:2000yn, Kallosh:2010ug} 
and thus we set $a_H = 0$ in this paper.%
\footnote{
In models of D-term inflation~\cite{Enqvist:1998pf, Enqvist:1999hv, Kawasaki:2001in},
the Hubble-induced mass as well as
the Hubble-induced $A$-term are absent: $c_{H} = a_{H} = 0$. 
However, the following discussion is also correct even if $c_H = 0$,
due to the Hubble friction. 
}

As we show below, 
we have to take into account 
higher-dimensional terms $V_K$ coming from a K$\ddot{\text{a}}$hler potential. 
Let us consider the following K$\ddot{\text{a}}$hler potential as an illustration for the origin of $V_K$:
\beq
K \sim I^\dagger I \frac{ \phi^{n'}}{\Mpl^{n'}},  
\eeq
where 
$I$ is a field which has a non-zero $F$-term during inflation (i.e. $ \abs{F_I}^2 = 3 H^2 \Mpl^2$). 
This operator gives the AD field a potential as%
\beq
V_K = \lmk 
\frac{-a_{H^2}}{n' \Mpl^{n'-2}} H^2 \phi^{n'} 
+ h.c. \rmk + \dots, 
\eeq
where "$\cdots$" denotes higher-dimensional Planck-suppressed terms. 
The parameter $a_{H^2}$ is an ${\cal O}(1)$ constant.

To sum up,
the potential of the AD field is given by the sum of 
the soft SUSY breaking terms $V_s$, 
the $F$-term potential $V_F$,
the Hubble-induced terms $V_H$, 
and the potential coming from a K$\ddot{\text{a}}$hler potential $V_K$: 
\beq
 V = V_{\text{s}} + V_F + V_{\text{H}} + V_K.
\eeq
In some cases, thermal potentials affect on the dynamics of the flat direction.
As we see below, however, 
we are interested in the case that the VEV of the AD field is 
so large to evade the baryonic isocurvature constraint, which results in a low reheating temperature.
Hence thermal potentials are irrelevant in the following discussion.
In this paper, 
we set all unknown ${\cal O}(1)$ parameters as one: 
$\abs{a_g}, \abs{c_H}, \abs{a_{H^2}} = 1$. 
We also assume $c_H < 0$, which makes 
the AD field obtain a large VEV during and after inflation as%
\footnote{
In the case of D-term inflation (i.e. $c_H = 0$), 
the AD field can stay anywhere $V'' \lesssim H$ is satisfied 
due to the Hubble friction. 
Eq.~(\ref{VEV}) is correct even in this case as long as the initial field value of the AD field is sufficiently large.
}
\beq
 \abs{\phi}  &\simeq& \lmk \frac{\abs{c_H}}{(n-1)}\rmk^{1/2(n-2)} 
\lmk \frac{1}{\lambda} H \Mpl^{n-3} \rmk^{1/(n-2)}, \\
&\simeq&
\left\{
\bea{ll}
1 \times 10^{16} \GEV \lambda^{-1/2} \lmk \frac{H}{10^{14} \GEV} \rmk^{1/2} 
\qquad \text{ for } n=4, \\
2 \times 10^{17} \GEV \lambda^{-1/4} \lmk \frac{H}{10^{14} \GEV} \rmk^{1/4} 
\qquad \text{ for } n=6, \\
4 \times 10^{17} \GEV \lambda^{-1/6} \lmk \frac{H}{10^{14} \GEV} \rmk^{1/6} 
\qquad \text{ for } n=8, 
\eea
\right.
\label{VEV}
\eeq
with the Hubble parameter being $H_{\text{inf}}$ and $H (t)$, respectively.
Note that if $\lambda = {\cal O}(10^{-4})$, the VEV of the AD field during inflation 
is as large as the Planck scale. 
Curvature of the phase direction of AD field, $\theta$, is dominantly given by $V_K$;
\beq
m_\theta^2 &\equiv& \frac{1}{2 \abs{ \phi}^2} \frac{\del^2 V}{\del \theta^2}, \\
&\simeq&
\frac{n' |a_{H^2}|}{2} H^2 \lmk \frac{ \abs{\phi}}{\Mpl} \rmk^{n'-2}. 
\label{m_theta}
\eeq
Note that the curvature is highly suppressed  compared with $H^2$ for $\abs{\phi} \lesssim \Mpl$, 
that is, for $\lambda \gtrsim 10^{-4}$.
(see Eqs.~(\ref{H_inf}) and (\ref{VEV})).

After inflation ends, the energy density of the Universe is dominated by coherently oscillating inflaton
and the Hubble parameter decreases as the Universe expands. 
When 
the Hubble-induced mass becomes less than 
the soft mass 
(i.e. $H (t) = H_{\text{osc}} \simeq m_\phi (\phi)$), 
the AD field begins to oscillate around the low-energy vacuum, $\phi=0$. 
Since the phase of the AD field just before the oscillation begins
is generally different from the 
one determined by the $A$-term of Eq.~(\ref{V_s}), 
the AD field also begins to rotate in the complex plane. 
Baryon number is generated by the rotation
because baryon density is given by
\beq
n_B 
 = -2b \Im \lkk \phi^* \dot{\phi} \rkk,
\eeq
where $b$ is the baryon charge of the AD field and the dot above the AD field denotes the derivative with respect to the time.
For example, $b = -1/3$ for $\bar{u} \bar{d} \bar{d}$ flat direction. 
The time evolution of the baryon density is written as
\beq
\dot{n}_B +3 H n_B =  2 b \Im \lkk \phi^* \frac{\del V}{\del \phi^*} \rkk.
\eeq
We estimate the solution of this equation as
\beq
\lmk \frac{a(t)}{a(t_{\text{osc}})} \rmk^3 n_B (t) 
&=& 2 b \int a^3 \Im \lkk \phi^* \frac{\del V}{\del \phi^*} \rkk \dd t , \\
&\sim& \lambda \frac{b}{H_{\text{osc}}} \frac{ \mg \abs{\phi_{\text{osc}}}^n}{\Mpl^{n-3}} \sin\lmk -n \theta \rmk, \label{n_b general}\\ 
&\sim&
\lmk b  \sin\lmk -n \theta \rmk \rmk \mg \abs{\phi_{\text{osc}}}^2,
\label{n_b}
\eeq
where $\theta$ is the initial phase of the AD field, 
which stays at a certain non-zero phase $\theta$ due to the Hubble friction before oscillation begins. 
We use $\abs{\phi_{\text{osc}}}^n \sim H_{\text{osc}} \abs{\phi_{\text{osc}}}^2 \Mpl^{n-3}/\lambda$ in the last line.
We define an ellipticity parameter as
\beq
 \epsilon &\equiv& \frac{n_B /\abs{b} }{n_\phi}, \label{ellipticity def}\\
&\sim& \text{sgn}(b) \sin\lmk - n \theta \rmk \frac{ \mg }{ H_{\text{osc}}}, \label{ellipticity}
\eeq
where $n_\phi \simeq \sqrt{V''} \abs{\phi}^2$ is the number density of the AD field. 
In this paper, we assume $\epsilon = \mg / H_{\text{osc}}$, 
because the initial phase of the AD field $\theta$ is of the order of unity unless it is fine-tuned 
to coincide with the low energy vacuum $\theta = 0$. 
Thus we obtain the present baryon-to-entropy ratio as
\beq
Y_B \equiv \frac{n_B}{s} &=& 
 \frac{3T_{\rm RH}}{4} \left.\frac{n_B}{\rho_{\rm rad}}\right|_{\rm RH} 
 = \frac{3T_{\rm RH}}{4} \left.\frac{n_B}{\rho_{\rm inf}}\right|_{\rm RH}, \\
&=&  \frac{3T_{\rm RH}}{4} \left.\frac{n_B}{\rho_{\rm inf}}\right|_{\rm osc}
\simeq  \frac{\epsilon \abs{b} T_{\rm RH}}{4 H_{\text{osc}}} \lmk \frac{ |\phi_{\text{osc}}|}{ \Mpl} \rmk^2, 
\label{Y_B}
\eeq
where we use $\rho_{\text{inf}}|_{\text{osc}} \simeq 3 \Mpl^2 H_{\text{osc}}^2$ in the last line.
Here we have assumed that
there is no entropy production other than the decay of inflaton,
which also implies that the AD field does not dominate the Universe.
Note that if the AD field releases its charge into the standard model particles 
before the sphaleron process freezes out, 
there is an ${\cal O}(1)$ correction to Eq.~(\ref{Y_B})~\cite{Fukugita:1986hr}. 
This is the case that we consider 
in Secs.~\ref{grav without Q} and \ref{GMSB without Q}.

\section{Baryonic isocurvature perturbation}
\label{sec:}

Since there is no sizable Hubble-induced A-term, the phase direction of the AD field 
develops quantum fluctuations during inflation, 
and as a result 
a baryonic isocurvature fluctuation, which is tightly constrained by recent observations,
is produced.
In this section, we consider the dynamics of the phase direction of the AD field in detail.

First we consider the case of $\lambda \gg 10^{-4}$, 
in which 
the VEV of the AD field is so small that the potential $V_K$ is negligible and 
the curvature of 
the phase direction is much less than the Hubble parameter during inflation (see Eq.~(\ref{m_theta})). 
The phase direction of the AD field therefore acquires quantum fluctuations during inflation as~\cite{Enqvist:1998pf, 
Enqvist:1999hv, Kawasaki:2001in, Kasuya:2008xp}
%
\beq
 \abs{\delta \theta} \simeq \frac{\sqrt{2}H_{\text{inf}}}{2 \pi \abs{\phi_{\text{inf}}}}. 
\eeq
Since the baryon number is related to the initial phase (see Eq.~(\ref{n_b})), 
this fluctuation induces a sizable baryonic isocurvature fluctuation 
as 
\beq
 \mathcal{S}_{b \gamma} \equiv \frac{\delta Y_B}{Y_B} \simeq n \cot \lmk n \theta \rmk \delta \theta.
\eeq

The baryonic isocurvature perturbation is constrained 
by observations of the cosmic microwave background, which have shown that
the density perturbations are predominantly adiabatic~\cite{Hinshaw:2012aka, Ade:2013zuv}. 
The $Planck$ Collaboration puts an upper bound on 
the totally uncorrelated isocurvature fraction 
as~\cite{Ade:2013uln}
\beq
 \frac{\mathcal{P_{SS}}(k_*)}{\mathcal{P_{RR}}(k_*) + \mathcal{P_{SS}}(k_*)} 
 \lesssim 0.039, 
\eeq
where $\mathcal{P_{RR}}$ and $\mathcal{P_{SS}}$ are power spectra of 
the adiabatic fluctuation and isocurvature fluctuation, respectively, 
and $k_*$ ($=0.05 \text{ Mpc}^{-1}$) is a pivot scale. 
Since we are interested in the baryonic isocurvature fluctuation, 
we use the following relation: 
\beq
 \mathcal{P_{SS}} = 
 \lmk \frac{\Omega_{\rm b}}{\Omega_{\rm DM}} \rmk^2 \mathcal{P}_{\mathcal{S}_{b \gamma} \mathcal{S}_{b \gamma}}, 
\eeq
where
$\Omega_{\rm b}$ and $\Omega_{\rm DM}$ are the density parameter
of the baryon and DM, respectively. 
Thus we obtain an upper bound on the baryonic isocurvature fluctuation as 
\beq
 \left\vert S_{{\rm b} \gamma} \right\vert \lesssim \frac{\Omega_{\rm DM}}{\Omega_{\rm b}} \lmk 0.039 \times 2.2 \times 10^{-9} \rmk^{1/2} 
 \simeq 5.0 \times 10^{-5}, 
\eeq
where we have used $\mathcal{P_{RR}}^{1/2} \simeq 2.2 \times 10^{-9}$~\cite{Ade:2013zuv}. 
The upper bound and the value of $H_{\text{inf}}$ indicated by BICEP2 put a lower bound on $\phi_{\text{inf}}$ as
\beq
 \abs{\phi_{\text{inf}}} \gtrsim 4 \times 10^{17} \GeV \times n \abs{\cot (n \theta)}. 
 \label{lower bound on VEV}
\eeq
%
This means that the VEV of the AD field should be as large as the Planck scale,
which implies that the higher-dimensional operator in the superpotential should be suppressed, 
$\lambda \lesssim {\cal O}(10^{-4})$ (see Eq.~(\ref{VEV})).
When the AD field obtains such a large VEV,
the potential $V_K$ is effective
and 
the phase direction of the AD field obtains 
a mass of the order of the Hubble parameter during inflation (see Eq.~(\ref{m_theta})).%
\footnote{
In contrast, if $\lambda = {\cal O}(1)$, 
the lower bound in Eq.~(\ref{lower bound on VEV}) requires 
about $1\%$ and $10\%$ tuning on the initial phase $\theta$ 
for $n=4$ and $n=6$ flat directions, respectively. 
This tuning would not be explained by the anthropic principle 
because human life would be able to exist whether or not baryonic isocurvature fluctuation exists. 
Taking this tuning seriously, we consider the case of $\lambda \lesssim 10^{-4}$ in this paper. 
} 
In this case, the baryonic isocurvature fluctuation is absent from the beginning.
To summarize,
in order to avoid a sizable baryonic isocurvature fluctuation, 
the VEV of the AD has to be as large as the Planck scale, 
in which case baryonic isocurvature fluctuation is absent 
due to the potential originated from a K$\ddot{\text{a}}$hler potential. 
Thus we assume 
$\lambda \lesssim 10^{-4}$ hereafter.%
\footnote{ 
The smallness of $\lambda$ might be understood by some flavor symmetry. 
}

Let us discuss the implication of the baryonic isocurvature fluctuation on
the reheating temperature using Eq.~(\ref{Y_B}). 
Since the VEV of the AD field at the beginning of its oscillation is 
related with that during inflation via $\abs{\phi_{\text{osc}}} = (H_{\text{osc}}/H_{\text{inf}})^{1/(n-2)} \abs{\phi_{\text{inf}}}$,
a large VEV during inflation results in a relatively large VEV at the onset of its oscillation,
and then the AD field tends to dominate the Universe.
Therefore,
in order to account for today's baryon-to-entropy ratio without additional entropy production
except for the decay of inflaton,
the reheating temperature tends to be small to dilute the AD field successfully.

From Eq.~(\ref{VEV}),
the VEV of the AD field at the beginning of its oscillation is given by
\beq
 \abs{\phi_{\text{osc}}} 
\simeq
 \left\{
 \bea{ll}
 4 \times 10^{12} \GEV \lmk \frac{\lambda}{10^{-4}} \rmk^{-1/2} 
 \lmk \frac{H_{\text{osc}}}{1 \TEV} \rmk^{1/2} 
 &\qquad \text{ for } n = 4, \\
 3 \times 10^{15} \GEV \lmk \frac{\lambda}{10^{-4}} \rmk^{-1/4} 
 \lmk \frac{H_{\text{osc}}}{1 \TEV} \rmk^{1/4} 
 &\qquad \text{ for } n = 6, \\
 3 \times 10^{16} \GEV \lmk \frac{\lambda}{10^{-4}} \rmk^{-1/6} 
 \lmk \frac{H_{\text{osc}}}{1 \TEV} \rmk^{1/6} 
 &\qquad \text{ for } n = 8, 
 \eea
 \right. 
 \label{VEV osc}
\eeq
where $H_{\text{osc}}$ is the Hubble parameter at the oscillation time. 
Thus the observed baryon density 
requires the reheating temperature of the Universe as 
\beq
 \TR \simeq 
  \left\{
  \bea{ll}
  4 \times 10^5 \GEV \  \epsilon^{-1} 
\lmk \frac{\lambda}{10^{-4}} \rmk 
  &\qquad \text{ for } n = 4, \\
   0.8 \GEV \  \epsilon^{-1} 
\lmk \frac{\lambda}{10^{-4}} \rmk^{1/2} 
  \left( \frac{H_{\text{osc}}}{1\text{ TeV}} \right)^{1/2} 
   &\qquad \text{ for } n = 6, \\ 
   9 \MEV \  \epsilon^{-1} 
\lmk \frac{\lambda}{10^{-4}} \rmk^{1/3} 
  \left( \frac{H_{\text{osc}}}{1\text{ TeV}} \right)^{2/3} 
   &\qquad \text{ for } n = 8, 
  \eea
  \right.
  \label{TR}
\eeq
where we have used $Y_B \simeq 8.7 \times 10^{-11}$ 
for the observed baryon-to-entropy ratio~\cite{pdg}, 
and assumed $b = - 1/3$. 
We should emphasize that the tight constraint on the baryonic isocurvature perturbation requires that $\lambda \lesssim 10^{-4}$ and
puts a severe upper bound on the reheating temperature,
if there is no additional entropy production.
Eq.~(\ref{TR}) is an important prediction 
from the BICEP2 result 
in a scenario for the Affleck-Dine baryogenesis. 
In Sec.~\ref{models}, 
we explain 
how 
the parameters $\epsilon$ and $H_{\text{osc}}$ is determined for each scenario 
in gravity and gauge mediation.

Here we comment on the gauge symmetry breaking 
and its effect on the reheating process~\cite{Allahverdi:2005mz}. 
Since 
the non-zero VEV of the AD field breaks gauge symmetries, 
one may consider that the reheating process might be hindered. 
The hindrance of the reheating process
occurs
if the AD field continues to oscillate coherently and if the Standard Model gauge symmetry 
is completely broken by the VEV of the AD field. 
The former condition is not satisfied in the case that Q-balls are formed, which we explain 
in the next section. 
The latter condition is not satisfied for $\bar{u} \bar{d} \bar{d}$ and $L H_u$ flat directions, 
for example. 
As explained in Sec.~\ref{models},
Q-balls are usually formed in gauge mediation 
except for the case of $L H_u$ flat direction. 
Thus, the suppression of the reheating process may be realized only in gravity mediation 
without Q-ball formation. 
This is the case we consider in Sec.~\ref{grav without Q}, 
in which we check that 
the reheating process is not affected by the AD field 
at least in the case we are interested in (see Eq.~(\ref{phi at TR})).

Finally, we comment on the case that 
the superpotential of the AD field is absent due to a discrete R-symmetry. 
In this case, the VEV of the AD field during inflation naturally becomes 
the Planck scale and thus a baryonic isocurvature fluctuation is absent. 
However, 
the VEV at the beginning of its oscillation is also the Planck scale, 
at which the AD field still feels $V_K$. 
Since the potential of the AD field $V_K$ is so complicated that 
the ellipticity parameter $\epsilon$ becomes ${\cal O}(1)$, 
Eq.~(\ref{Y_B}) implies that the reheating temperature has to be much lower than $1 \MEV$ in this case, 
which spoils the success of the BBN scenario. 
Thus we consider the case of $W \ne 0$.

\section{\label{Q-ball}Q-balls}

In this section, we explain the dynamics of the Q-ball, which is a non-topological soliton 
formed after the Affleck-Dine mechanism 
in many SUSY models~\cite{Coleman, Qsusy, KuSh, EnMc, KK1, KK2, KK3}. 
We first explain the condition of the Q-ball to be formed and then review decay processes of the Q-ball.

\subsection{Formation of Q-ball}
After the AD field starts to oscillate and rotate around the low energy vacuum,
the amplitude of the oscillation decreases due to the Hubble expansion. 
Since baryon number-violating terms are higher-dimensional ones, 
their effects become irrelevant 
and the generated baryon number is conserved 
soon after the beginning of the oscillation. 
Thus, in this section,
we assume baryon number to be conserved
and investigate the stable configuration of the AD field 
in a system with non-zero baryon charge.

The energy of the AD field is given as
\begin{eqnarray}
E = \int  \dd^3x \lkk \dot{\abs{\phi}}^2 + \abs{\nabla \phi}^2 + V \left( \abs{\phi} \right) \rkk.
\label{E}
\end{eqnarray}
We are interested in the case with sufficiently small value of the AD field, for which the potential is approximated by $V = m_\phi^2 (\phi) \abs{\phi}^2$. 
Since 
the baryon density is already produced by the Affleck-Dine mechanism, 
we consider a system with non-zero baryon charge. 
The baryon charge is given by
\begin{eqnarray}
 Q = - 2 \int  \dd^3x \Im \lkk 
 \phi^* \dot{\phi} \rkk, 
\end{eqnarray}
where we have omitted the factor $b$ for notational simplicity.
The scalar field configuration which minimizes the energy given in Eq.~(\ref{E}) 
with a fixed baryon charge $Q$ is obtained by minimizing 
the following combination;
\begin{equation}
E + \omega_0 \left[ Q + 2 \int \dd^3 x \Im \lkk \phi^* \dot{ \phi} \rkk \right], 
\end{equation}
where $\omega_0$ is a Lagrange multiplier. 
Terms with time derivatives are rewritten as 
\beq
 \dot{\abs{\phi}}^2 + 2 \omega_0 \Im \lkk \phi^* \dot{\phi} \rkk = 
 \abs{\dot{\phi} + i \omega_0 \phi}^2 - \omega_0^2 \abs{\phi}^2.
\eeq
The minimization condition determines the time dependence of the AD field as 
\beq
 \phi(\bm{r},t)= \varphi(\bm{r}) e^{-i \omega_0 t} / \sqrt{2}. 
\eeq
Assuming a spherically symmetric ansatz, $\varphi( \bm{r})=\varphi(r )$, 
we obtain the following equation which determines $\varphi(r)$:
\begin{equation}
\frac{\partial^2}{\partial r^2} \varphi + \frac{2}{r} \frac{\partial}{\partial r} \varphi + \omega_0^2 \varphi - 
\frac{\partial}{\partial \varphi} V ( \varphi ) = 0.
\label{Q-ball eq}
\end{equation}
The boundary condition is $\varphi'(0 )=0$ and $\varphi (\infty) = 0$ 
since 
we are interested in smooth and localized configurations. 
Regarding $\phi$ and $r$ as a position and a time variable, 
Eq.~(\ref{Q-ball eq}) is interpreted as the equation of motion of a particle in one dimension
with a friction term $(2/r) \del \varphi / \del r$. 
Using this analogy, one can find the following condition for existence of 
a spatially localized configuration, referred to as Q-ball~\cite{Coleman}: 
\beq
\text{min}_\varphi \lkk \frac{ 2 V(\varphi)}{  \varphi^2 } \rkk < \omega_0^2 < \frac{\del^2 V(0)}{\del \varphi^2}. 
\label{Q condition}
\eeq
In most cases we are interested in, the energy of the Q-ball per unit charge is well approximated by $\omega_0$.

Linear analyses 
indicate that 
there are instability bands during the oscillation of the AD field which corresponds to a typical size of the Q-ball $R$ 
if the condition of Q-ball formation Eq.~(\ref{Q condition}) is satisfied~\cite{KuSh, EnMc}.
This means that 
the coherently oscillating AD field is unstable and fragments into Q-balls 
soon after the onset of its oscillation. 
A typical charge of Q-balls is roughly estimated by 
the charge which is contained in the volume of a typical Q-ball size $R$ at the formation time:
\beq
 Q \sim n_B(t_{form}) R^{3} \sim \lmk \frac{a(t_{\text{osc}})}{a(t_{form})} 
 \rmk^3 \omega_0 \abs{\phi_{\text{osc}}}^2 R^3,
 \label{Q charge}
\eeq
where we have included the dependence on the scale factors $a$ 
because it needs some time for Q-ball to be formed completely. 
The numerical simulations have shown that 
Q-balls are indeed formed when the condition of Q-ball formation Eq.~(\ref{Q condition}) is satisfied, 
and have determined the proportional constant including 
$a^3(t_{\text{osc}}) / a^3(t_\text{form})$ of Eq.~(\ref{Q charge}) 
for the cases we are interested in (see Eqs.~(\ref{Q in grav 0}) and (\ref{Q in gauge 0}))~\cite{KK1, KK2, KK3, Hiramatsu:2010dx}.

\subsection{Decay of Q-ball}
To explain the decay of Q-balls, 
let us focus on 
a Q-ball which consists only of squarks. 
Numerical simulations have shown that 
almost all of the baryon charge of the AD field
are transferred into 
Q-balls~\cite{KK1, KK2, KK3, Hiramatsu:2010dx}. 
Then Q-balls decay and release their baryon charge into quarks if they are unstable. 
The AD field interacts with quarks via gauge interactions 
and thus Q-balls lose their baryon charge by emitting quarks from their surfaces~\cite{evap}.%
\footnote{Far inside Q-balls, field values of squarks are large and hence gauginos and quarks are heavy. Therefore, Q-balls cannot decay into them.}
The condition for Q-ball decay is that the energy of the Q-ball per unit baryon charge, 
$\omega_0/\abs{b}$, 
is larger than masses of baryons in the hadron phase, $m_b \simeq 1 \GeV$.

Usually, the baryon charge density inside a Q-ball is so large that 
a naive rate estimated by squark decay 
exceeds an upper limit 
by the Pauli blocking effect. 
The rate of particle emission from a Q-ball surface is therefore 
determined by the Pauli blocking effect on its surface and is given as~\cite{evap}%
\footnote{
Here we assume that 
quarks and gauginos are massless. 
A correction for the flux of massive particles is derived in the Appendix. 
}
\beq
 \frac{dN}{dt} &\simeq& \sum_i 4 \pi \tilde{R}^2 \bm{n \cdot j_i}, \\
 \bm{n \cdot j_i} &\simeq& 2 \int \frac{\dd^3 k }{(2 \pi)^3} \theta \lmk E_i/2 - 
 \abs{\bm{k}} \rmk \theta \lmk \bm{k \cdot n} \rmk \bm{\hat{k} \cdot n} \nonumber \\
 &=& \frac{E_i^3}{96 \pi^2}, 
 \label{flux}
\eeq
where $n$ is the outward-pointing normal vector, $\bm{j}$ is particle flux, and 
$\tilde{R}$ is the effective radius of the Q-ball
given by $\phi(\tilde{R}) \sim \omega_0$~\cite{KY}. 
The interaction energy 
$E_i$ is given by the energy of the Q-ball per unit charge, $\omega_0$, 
when the relevant elementary process is squark decay, such as $\tilde{q} \to q + (\text{gaugino})$. 
In addition, 
the baryon charge density 
inside the Q-ball is so large that 
the scattering process via gaugino (or Higgsino) exchange $\tilde{q} + \tilde{q} \to q + q$ occurs
efficiently. 
It has been shown that 
the rate of this process is also saturated by the Pauli blocking effect, and 
the interaction energy $E_i$ is given by $2 \omega_0$ in this case~\cite{KY}. 
The rate of Q-ball decay is dominated by the latter process and thus its lifetime $\Gamma_Q^{-1}$ 
is given as
\beq
 \Gamma_Q^{-1} &\simeq& \lmk \frac{1}{Q} \frac{\dd N}{\dd t} \rmk^{-1}, \\
 &\simeq& \lmk 8 n_q \frac{\tilde{R}^2 \omega_0^3}{24 \pi Q} \rmk^{-1}, 
 \label{Q decay}
\eeq
where $n_q$ is the number of species for quarks interacting with the AD field and is typically ${\cal O}(10)$.

Q-balls completely lose their charge and energy 
when the condition $\Gamma_Q \sim H$ is satisfied. 
The decay temperature of Q-ball is thus determined as
\beq
 \Td \simeq 
  \left\{
  \bea{ll}
 \TR 
 \lmk \sqrt{\frac{30}{\pi^2 g_*} } \frac{\Gamma_Q \Mpl}{\TR^2} \rmk^{1/4}
  &\qquad \text{ for } \Td > \TR, \\
 \lmk \frac{90}{4 \pi^2 g_* } \rmk^{1/4} \sqrt{\Gamma_Q \Mpl} 
   &\qquad \text{ for } \Td < \TR, 
  \eea
  \right.
  \label{Tdecay}
\eeq
where 
the first line is 
the case where 
Q-balls decay before reheating completes 
while the second one is 
the case where
Q-balls decay after reheating completes. 
In the latter case, 
the energy density of Q-balls may dominate the Universe. 
Since Q-balls are localized lumps much smaller than the horizon scale, 
their energy density 
decreases as $a^{-3}$, 
where $a$ is a scale factor. 
Thus the energy density of Q-balls dominates that of the Universe 
when 
the following condition is satisfied:
\beq
 1 
 &\lesssim&
 \left. \frac{\rho_Q}{\rho_r} \right\vert_{T = \Td} 
 \simeq
 \left. \frac{\rho_Q}{\rho_r} \right\vert_{T = \TR} \lmk \frac{\TR}{\Td} \rmk, \\
 &\simeq&
 \left. \frac{\rho_\phi}{\rho_I} \right\vert_{\text{osc}} \lmk \frac{\TR}{\Td} \rmk 
 \simeq
 \frac{\phi_{\text{osc}}^2}{3 \Mpl^2} \lmk \frac{\TR}{\Td} \rmk, 
 \label{non dominant cond.}
\eeq
where $\rho_I$ is the inflaton energy. 
We checked that the energy density of Q-balls never dominate that of the Universe 
in the case we consider.

As explained above, 
Q-balls dominantly decay into quarks. 
However, 
they decay into SUSY particles 
if the decay process is kinematically allowed. 
From kinematics and the conservation of baryon charge, 
Q-ball can decay only into particles 
lighter than the energy of the Q-ball per unit charge, $\omega_0$. 
Since 
$\omega_0$ 
is less than the mass of the AD field (see Eq.~(\ref{Q condition})), 
Q-balls cannot decay into the AD field itself. 
This is another explanation of the stability of Q-ball. 
On the other hand, 
Q-balls decay into gauginos and/or Higgsinos 
if they interact with the Q-balls and their masses are less than $\omega_0$. 
However, 
in contrast to the case of quarks, 
gauginos and Higgsinos cannot be produced through 
a scattering process like $\tilde{q} + \tilde{q} \to (\text{gaugino}) + (\text{gaugino})$
due to the conservation of baryon charge. 
Thus 
if we could neglect their masses, 
their production rate from Q-ball decay is 
given by Eq.~(\ref{flux}) with $E_i = \omega_0$. 
In addition, there is a correction coming from non-zero masses of gauginos and Higgsinos 
as explained in the Appendix.

\section{\label{models}scenarios for dark matter production in low reheating temperature}

To construct a consistent cosmological scenario, 
it is necessary to account for the following observed DM density as well as the baryon density: 
\beq
 \frac{\rho_{\text{DM}}}{s}  &\simeq& 
 (3.5 \EV) \times \Omega_{\text{DM}} h^2, \\
  &\simeq& 
 0.44 \EV, 
 \label{rhoDM}
\eeq
where we have used $\Omega_{\text{DM}} h^2 \simeq 0.12$ for the observed DM abundance 
in the last line~\cite{pdg}. 
As explained in Sec.~\ref{sec:}, 
the reheating temperature of the Universe must satisfy
Eq.~(\ref{TR})
to account for the observed baryon density by the Affleck-Dine baryogenesis. 
In this section, 
we 
propose scenarios to account for the DM density 
in that reheating temperature.

Assuming models of gravity mediation and gauge mediation, 
we investigate scenarios in each model with and without Q-ball formation. 
In Sec.~\ref{grav without Q}, 
we consider 
a model of gravity mediation without Q-ball formation, in which
DM is mainly produced from two processes;
decay and inelastic scatterings during reheating era~\cite{Allahverdi:2002nb, HKMY}. 
In Sec.~\ref{grav with Q}, 
we consider a model of gravity mediation with Q-ball formation.
In this case, DM can be produced from Q-balls~\cite{EnMc, 
Fujii:2001xp, Fujii:2002kr, Fujii:2002aj, Roszkowski:2006kw, Seto:2007ym, KKY} 
as well as by the above two processes.
In Sec.~\ref{GMSB without Q},
we consider a model of gauge mediation without Q-ball formation, 
which is only the case for $L H_u$ flat direction. 
In this case, gravitino DM is produced from the thermal plasma 
at the time of reheating~\cite{Moroi:1993mb}
as well as from decay and inelastic scatterings during reheating era. 
A model of gauge mediation with Q-ball formation is investigated in Sec.~\ref{GMSB with Q}. 
Also in this case, Q-balls provide another source of DM~\cite{ShKu, 
Doddato:2011fz, Kasuya:2011ix, Doddato:2012ja, Kasuya:2012mh}. 

\subsection{\label{grav without Q}Gravity mediation without Q-ball
}

In models of gravity mediation, 
the mass of the AD field $m_\phi (\abs{\phi})$ logarithmically depends on $\abs{\phi}$ 
due to renormalization group running of squark masses, 
and the condition for Q-ball formation, Eq.~(\ref{Q condition}), is satisfied 
when the mass decreases with increasing the VEV of the AD field 
(i.e. $\dd m_\phi / \dd \phi < 0$)~\cite{EnMc}. 
While the strong interaction makes
squarks light with increasing energy scale, 
Yukawa interactions make them heavy. 
In a typical model of gravity mediation, 
stops become heavy with increasing energy scale
while the first and second family squarks become light. 
Thus 
there may be no Q-ball solution 
if the AD field consists mainly of third family squarks. 
In this subsection, 
we consider the case without Q-ball formation in gravity mediation.

Since gravitino mass is of the same order as squark masses in gravity mediation, 
the ellipticity parameter, $\epsilon$ ($= \mg/ H_{\text{osc}} \simeq \mg / m_\phi(\phi_{\text{osc}})$), 
is of the order of one (see Eq.~(\ref{ellipticity})). 
Thus we obtain the VEV of the AD field at the beginning of its oscillation 
and the reheating temperature of the Universe from Eqs.~(\ref{VEV osc}) and (\ref{TR}). 
Assuming that the AD field continues to oscillate until reheating completes, 
we calculate the amplitude of the oscillation at the time of reheating as 
\beq
 \left. \phi \right\vert_{T= \TR} 
 &\simeq& 
 \phi_{\text{osc}} \frac{ \TR^2}{H_\text{osc} \Mpl}, \\
 &\simeq& 
 \left\{ 
 \bea{ll}
 3 \times 10^2 \GEV 
 \lmk \frac{\lambda}{10^{-4}} \rmk^{3/2}
 \lmk \frac{H_\text{osc}}{1 \TEV} \rmk^{-1/2}
 &\qquad \text{ for } n=4, \\
 7 \times 10^{-7} \GEV 
 \lmk \frac{\lambda}{10^{-4}} \rmk^{3/4}
 \lmk \frac{H_\text{osc}}{1 \TEV} \rmk^{1/4}
 &\qquad \text{ for } n=6, \\
 9 \times 10^{-10} \GEV 
 \lmk \frac{\lambda}{10^{-4}} \rmk^{1/2}
 \lmk \frac{H_\text{osc}}{1 \TEV} \rmk^{1/2}
 &\qquad \text{ for } n=8. 
 \eea
 \right.
 \label{phi at TR}
\eeq
Thus we find that the amplitude of the oscillation at the time of reheating 
is much less than the reheating temperature. 
This indicates that the AD field cannot affect the reheating process 
even if it could continue to oscillate coherently. 
Of course, the coherent oscillation dissipates through interactions with the thermal plasma 
much before reheating completes, 
but in this paper we do not investigate this issue in detail 
(see Ref.~\cite{Allahverdi:2005mz, Mukaida:2012qn} and references therein).

The Affleck-Dine baryogenesis with $n=4$ flat direction 
predicts the reheating temperature as $\TR \simeq 4 \times 10^5 \GEV$
for $\lambda = 10^{-4}$, 
in which case 
DM is produced thermally. 
Since the LSP is mostly bino-like 
in a typical model of gravity mediation, 
thermally produced binos
over-closes the Universe 
unless the mass of bino is fine-tuned to co-annihilate~\cite{Griest:1990kh,Gondolo:1990dk} 
with the stau~\cite{Ellis:1998kh}. 
On the other hand, 
when we consider $n=4$ flat direction with $\lambda \lesssim 10^{-10}$ or $n=6$, $8$ flat direction
with $\lambda \lesssim 10^{-4}$, 
the required reheating temperature is less than $1 \GEV$
(see Eq.~(\ref{TR})). 
Such a low reheating temperature 
results in a non-thermal production of DM, 
which we have investigated in Ref.~\cite{HKMY}. 
Here we review this calculation and apply it in our situation. 
In a low reheating temperature, 
DM is produced mainly through two contributions: 
direct decay of inflaton into SUSY particles~\cite{Moroi:1994rs, Kawasaki:1995cy, 
Moroi:1999zb, Allahverdi:2002nb, Gelmini:2006pw, Kurata:2012nf} and 
inelastic scatterings between high energy particles 
and thermal plasma during reheating process~\cite{Allahverdi:2002nb, HKMY}.%
\footnote{
DM is also produced thermally if the maximal temperature of the Universe is larger than 
the freeze-out temperature of the LSP~\cite{Chung:1998rq, Giudice:2000ex}. 
However, this contribution is always subdominant as shown in Ref.~\cite{HKMY}, 
and we neglect it in this paper. 
}

DM can be produced directly by the decay of inflaton~\cite{Moroi:1994rs, 
Kawasaki:1995cy, Moroi:1999zb, Gelmini:2006pw, Kurata:2012nf}
and its abundance is calculated as 
\beq
 \frac{\rho_{\text{DM}}^{\text{dir}}}{s} 
 &\simeq& m_{\text{DM}}
 \left. \frac{3 \TR n_{\text{DM}}}{4 \rho_{I}} \right\vert_{T = \TR}, \\
 &\simeq& m_{\text{DM}} \frac{3 \TR}{4 m_I} \Br_I, \\
 &\simeq& 0.4 \EV 
 \lmk \frac{m_{\text{DM}}}{1 \TeV} \rmk
 \lmk \frac{\TR}{ 0.1 \GeV} \rmk
 \lmk \frac{m_I}{10^{12} \GeV} \rmk^{-1}
 \lmk \frac{\Br_I}{5} \rmk,
 \label{rhoDM dir}
\eeq
where $m_{\text{DM}}$ is the mass of the LSP, 
and $m_I$ is the mass of the inflaton. 
We write $\Br_I$ as 
the average number of DM produced by each inflaton decay. 
One may expect $\Br_I = 1$ due to SUSY. 
In Ref.~\cite{Kurata:2012nf}, however, it was pointed out that 
SUSY particles are produced through a cascade shower from inflaton decay, 
like pions in a QCD shower, 
and the effective branching ratio $\Br_I$ deviates from $1$. 
They calculated $\Br_I$ and found that 
$\Br_I$ is as large as 
${\cal O}(10^{0-1})$ for $m_I = 10^{10} \GeV$. 
They also extrapolated the results from $m_I \le 10^{10} \GEV$ to larger inflaton mass 
and estimated $\Br_I = {\cal O}(10^{0-2})$ for $m_I = 10^{13} \GeV$.

Next, let us explain the other process of DM production; 
inelastic scattering process during the reheating era~\cite{Allahverdi:2002nb, HKMY}. 
At the first stage of reheating, 
the inflaton decays into high energy particles with 
the energy of the order of the inflaton mass $\mphi$. 
The daughter particles 
interact with each other and produce many low energy particles almost without losing their energy. 
Then low energy particles thermalize by their own interactions, 
but the energy of the thermal plasma is still much smaller than that of the inflaton. 
Once the thermal plasma is created in the Universe, 
the high energy particles produced from the inflaton decay inelastically interact with 
the thermal plasma, through which the number of high energy particles 
drastically increases~\cite{Kurkela:2011ti,Harigaya:2013vwa} 
(see also Ref.~\cite{Davidson:2000er, Arnold:2002zm,Jaikumar:2002iq}). 
Throughout this process, 
DM is produced in a certain amount as we explain below.%
\footnote{
The following estimation of the DM abundance involves uncertainties.
For example, the estimation of the rate of inelastic scatterings is based on qualitative discussions.
Therefore, the prediction on the DM mass given below has 
an uncertainty within about one order of magnitude. 
}

Let us estimate an inelastic scattering cross section between
a high energy particle and the thermal bath. 
It is enhanced by $t$-channel contribution and is given as
\beq
 \sigma_{\rm inelastic} 
 \sim \frac{\alpha^3}{t} 
 \sim \frac{\alpha^2}{T^2},
\eeq
where an infrared divergence is naturally regulated by the thermal mass of an internal field,
$\alpha^{1/2} T$. 
However,
in estimating the scattering rate, it is necessary to
take into account an interference effect between a daughter particle and 
its parent particle, 
known as the Landau-Pomeranchuk-Migdal (LPM) effect~\cite{Landau-Pomeranchuk, Migdal,
Gyulassy:1993hr,
Arnold:2001ba, Arnold:2001ms, Arnold:2002ja, Besak:2010fb}.
The interference effect forbids subsequent scattering processes 
while their phase spaces overlap with each other. 
When we write the position vector of a parent particle as $x^\mu = \lmk \Delta t, \Delta t \hat{z} \rmk$,
the interference effect
remains until the phase factor varies significantly as%
\beq
 1 \lesssim
 k x \sim \Delta t k^0 \theta^2 \sim \Delta t k^2_{\perp} / k^0,
 \label{delta t}
\eeq
where
$k$ and $k_\perp$ are the four momentum and 
the perpendicular momentum of the daughter particle, respectively.\footnote{
	Here and hereafter we assume that the daughter particle is 
	charged under a non-Abelian gauge group.
}
We write 
the emission angle of the daughter particle as 
$\theta$ ($=k_\perp / k^0$). 
From Eq.~(\ref{delta t}), 
we find that the LPM effect suppresses subsequent inelastic scattering processes 
during the interval $\Delta t (k^0) \sim k^0 / k_\perp^2$. 
We therefore determine the inelastic scattering rate 
as 
\beq
 \Gamma_{\rm inelastic} 
 \sim 
\min \lkk \la \sigma_{\rm inelastic} n \ra, 
 \int \frac{\dd k^0}{k^0} \frac{\alpha}{\Delta t (k^0)}, 
 \rkk. 
 \label{inela rate}
\eeq

Let us estimate the perpendicular momentum of the daughter particle $\Delta k_{\perp}$. 
The perpendicular momentum evolves as random walk due to elastic scatterings with the thermal plasma 
as
\beq
 \lmk \Delta k_\perp \rmk^2 \sim \qel t,
 \label{delta k}
\eeq
where $\qel$ is a diffusion constant given 
by the elastic scattering rate $\Gamma_{\text{el}}$ ($\simeq \alpha T$) as 
\beq
 \qel \sim \int \dd^2 q_\perp \frac{\del \Gamma_{\rm el}}{\del q_\perp^2} q_\perp^2 \sim \alpha^2 T^3. 
 \label{qel}
\eeq
Using Eqs.~(\ref{delta k}) and (\ref{qel}), we obtain
\beq
 \Delta t \sim \lmk \frac{k^0}{\qel} \rmk^{1/2} \sim \frac{1}{\alpha T} \lmk \frac{k^0}{T} \rmk^{1/2}.
 \label{del t 2}
\eeq

The rate of inelastic scatterings is determined by Eqs.~(\ref{inela rate}) and (\ref{del t 2}), 
from which we find that the energy loss rate increases with increasing the energy of the daughter particle. 
Therefore 
high energy particles from the inflaton decay continue to split into high energy particles 
whose energy is less than but the same in order of magnitude as parent particle's energy. 
Throughout this splitting process, 
the number density of high energy particles 
increases exponentially~\cite{Kurkela:2011ti,Harigaya:2013vwa}. 
Given a certain time when their energy is of the order of $E$, 
we can estimate 
their number density $n_h$ from the energy conservation as 
\beq
\label{number high}
 n_h \sim \frac{\mphi}{E} n_{\rm I}, 
\eeq
where $n_{\rm I}$ is the number density of the inflaton~\cite{HKMY}.

During the above splitting process, 
inelastic scatterings between a high energy particle and the thermal plasma  can produce DM, 
when the center-of-mass energy $\sqrt{4 T E}$ is 
larger than the mass of DM $\mDM$. 
The production rate 
is given as~\cite{Allahverdi:2002nb}
\beq
  \Gamma_{\rm DM} \sim \la \sigma_{\rm DM} n_r \ra \sim \frac{\alpha^2 T^3}{\mDM^2}, 
\label{Gamma_DM}
\eeq
for a reaction whose center-of-mass energy is just above the threshold 
(i.e. $E \gtrsim E_{\text{th}} \equiv \mDM^2/4 T$). 
Although the fine structure constant $\alpha$ for the DM production process
is in general different from the one for the inelastic scattering process, 
we neglect the difference in this paper for simplicity. 
The energy density of DM 
which is produced during the splitting process of high energy particles 
is therefore obtained as
\beq
\label{entropy ratio}
 \frac{\rho_{\rm DM}^{{\rm inela}}}{s} 
 &\sim& \mDM \frac{\Gamma_{\rm DM}}{\Gamma_{\rm inelastic}} \frac{n_h}{s}, \\ 
 &\sim& 
 \mDM 
 \frac{\alpha^2 \TR^3}{\mDM^2} \frac{\sqrt{\Eth}}{\alpha^2 \TR \sqrt{\TR}} 
 \frac{\mphi}{\Eth}
 \frac{n_{\rm I}}{s} 
 \\
 &\sim& 
 \frac{\TR^3}{\mDM^2}, \\
 &\sim& 
 1 \EV \lmk \frac{\TR}{0.1 \GEV} \rmk^3 
 \lmk \frac{\mDM}{1 \TEV} \rmk^{-2}.  
 \label{main result}
\eeq
Note that 
the DM abundance is independent of the mass of the inflaton.
In general, the gauge coupling constants in the second line cannot be 
canceled with each other, 
but our conclusion is still correct with an uncertainty within one order of magnitude.

Here we comment on some assumptions which we have implicitly used in the above calculation. 
We assume that 
the inflaton is so heavy that its daughter particles can produce DM 
at the time of $T=\TR$ (i.e. $\mphi \ge \mDM^2 / 2\TR \sim 10^{7} \GEV$). 
Since the energy scale of inflation is large, it is expected that the inflaton mass is also large.
The other case has been calculated in the original paper~\cite{HKMY}. 
In addition, the result in Eq.~(\ref{main result}) is over-estimated 
for the reheating temperature smaller than the QCD scale ($\sim 0.1 \GEV$), 
because some hadrons, such as neutral pion, have no gauge interactions 
and the energy loss by splitting processes 
may be suppressed.

Finally, let us consider 
the annihilation of DM. 
The DM abundance is reduced by its annihilation 
if $\la \sigma_{\text{ann}} v \ra n_{\text{DM}} \gtrsim H_{T = \TR}$, 
where $\la \sigma_{\text{ann}} v \ra$ is the thermally averaged annihilation cross section of DM. 
This implies that there is an upper bound on the DM abundance given as 
\beq
 \label{rho_anni}
 \frac{\rho_{\text{DM}}^{\text{ann}}}{s}
 &\simeq& 
 \sqrt{\frac{45}{8 \pi^2 g_*}} \frac{\mDM}{\TR \la \sigma_\text{ann} v \ra \Mpl}, \nonumber\\
 &\simeq& 
 0.9 \EV 
 \lmk \frac{\TR}{1 \GEV} \rmk^{-1} 
 \lmk \frac{\mDM}{1 \TEV} \rmk
 \lmk \frac{\la \sigma_\text{ann} v \ra  }{ 0.1   \TEV^{-2} } \rmk^{-1}. 
\eeq
The LSP is mostly bino-like in a typical model of gravity mediation 
Its annihilation effect is negligible as far as 
its abundance is consistent with the observed DM abundance.

\begin{figure}[t]
\centering 
\includegraphics[width=.45\textwidth, bb=0 0 450 455]{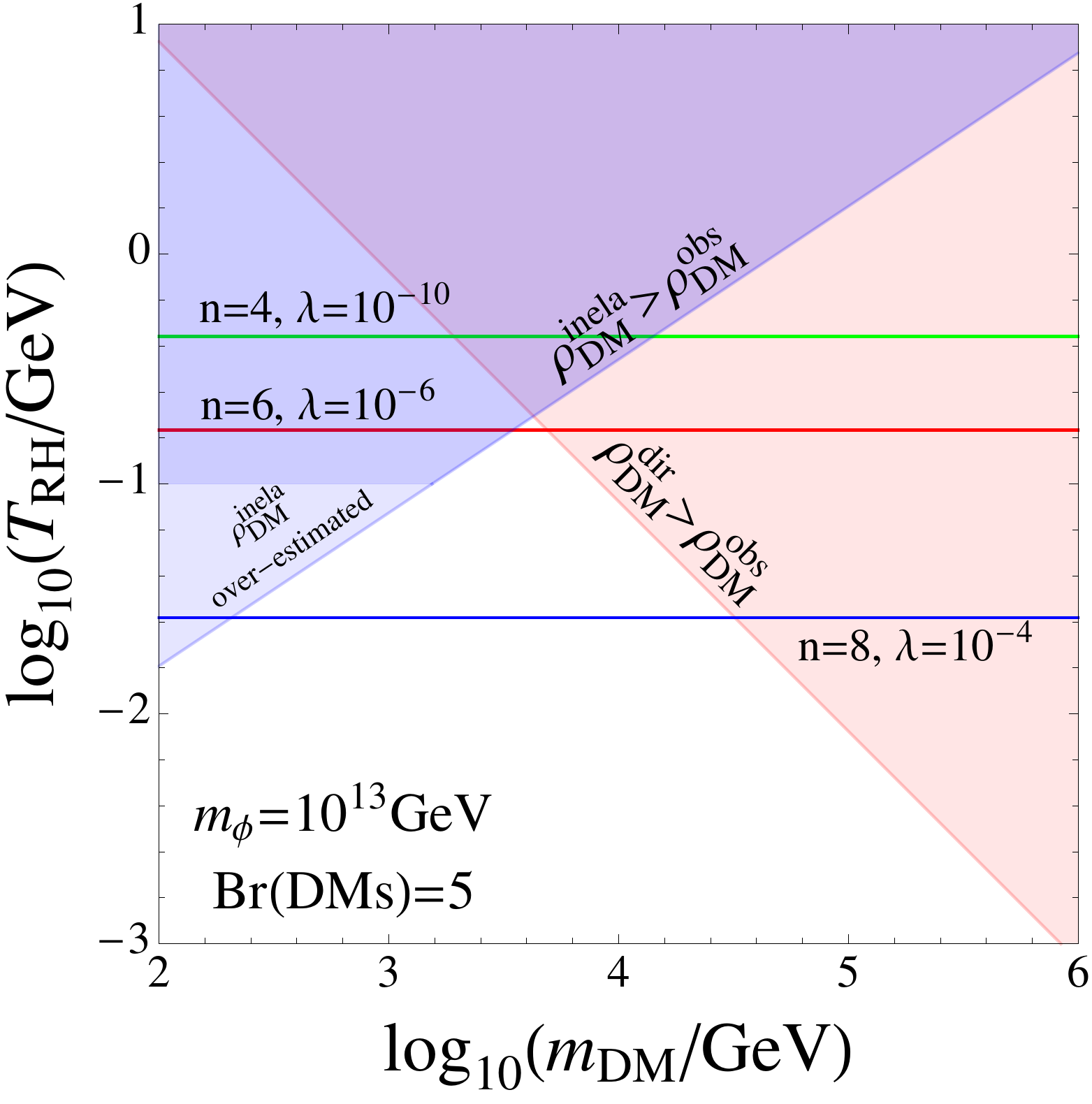} 
\hfill
\includegraphics[width=.45\textwidth, bb=0 0 450 455]{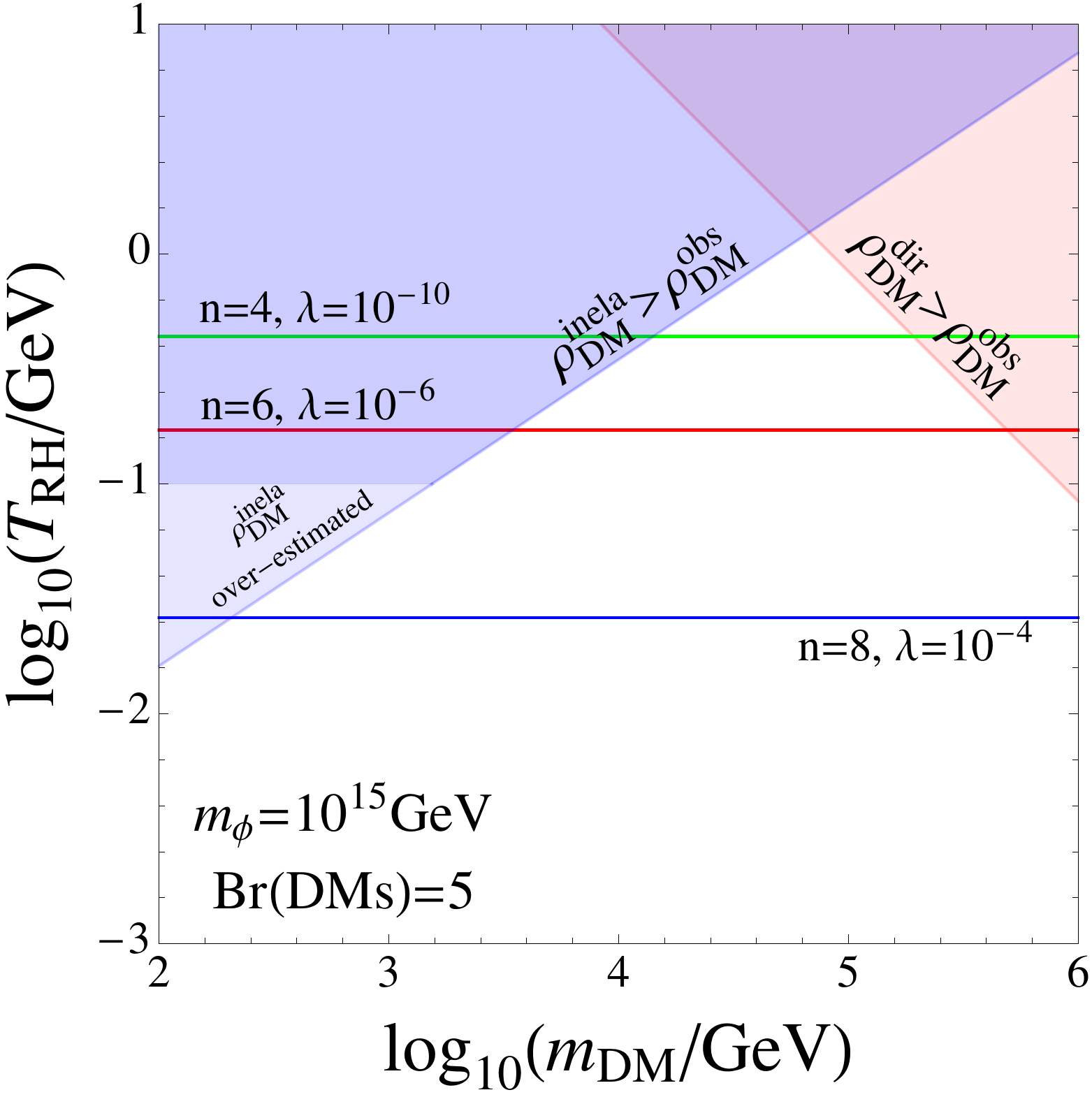} 
\caption{
Exclusion plot in a model of gravity mediation without Q-ball formation. 
We assume that 
the effective branching of inflaton decay into DM is $5$ and 
the mass of the inflaton $\mphi$ is $10^{13} \GEV$ (left panel) and $10^{15} \GEV$ (right panel). 
The abundance of DM produced from
direct decay of inflaton ($\rho_{\rm DM}^{\rm dir}$) 
and inelastic scatterings ($\rho_{\rm DM}^{\rm inela}$)
is larger than that observed in the red (light gray) and blue (dark gray) shaded region, respectively. 
The blue (dark gray) shaded region with a temperature lower than $0.1 \GEV$ is 
the region in which $\rho_{\rm DM}^{\rm inela}$ is over-estimated. 
The horizontal green-dashed, red-dotted, and blue lines are 
the reheating temperature required by the success of the Affleck-Dine baryogenesis 
for $n=4$ with $\lambda = 10^{-10}$, $n=6$ with $\lambda = 10^{-6}$, 
and $n=8$ with $\lambda = 10^{-4}$, respectively. 
We take $H_\text{osc} = 5 \TEV$. 
}
  \label{fig1}
\end{figure}

In summary, 
DM can be produced from two sources, 
and its abundance is given by the sum of Eqs.~(\ref{rhoDM dir}) and (\ref{main result}). 
Taking the annihilation of DM into account, 
we conclude that 
the DM abundance is given by 
\beq
 \frac{\rho_{\text{DM}}}{s} \simeq 
 \text{min} \lkk 
 \frac{\rho_{\text{DM}}^{\text{dir}}}{s} + \frac{\rho_{\text{DM}}^{\text{inela}}}{s}, 
 \frac{\rho_{\text{DM}}^{\text{ann}}}{s} \rkk. 
\eeq
In Fig.~\ref{fig1}, we show the constraint on the DM mass $m_{\rm DM}$ and the reheating temperature $T_{\rm RH}$.
The left and the right panels assume the inflaton mass $m_{\phi}$ of $10^{13}$ GeV and $10^{15}$ GeV, respectively.%
\footnote{In large field models discussed in Refs.~\cite{Takahashi:2010ky,Harigaya:2012pg}, the inflaton mass can be as large as $10^{15}$ GeV.}
On the boundary of and inside the red (light gray) shaded region, 
the decay of the inflaton produces DM density equal to and 
larger than the observed one.
On the boundary and inside the blue (dark gray) shaded region, 
inelastic scatterings during the thermalization process produce the correct and larger amount of DM.
As we have mentioned, for $T_{\rm RH}\lesssim 0.1~{\rm GeV}$, the thermalization process involves scatterings of hadrons and the DM abundance produced by inelastic scatterings 
may be over-estimated.
We also show the requirement of the reheating temperature from the successful Affleck-Dine Baryogenesis.
The green-dashed, red-dotted, and blue lines show
the required reheating temperature
for $n=4$ with $\lambda = 10^{-10}$, $n=6$ with $\lambda = 10^{-6}$, 
and $n=8$ with $\lambda = 10^{-4}$, respectively.

From Fig.~\ref{fig1},
we predict the mass of DM as 
\beq
 \mDM \simeq 
 \left\{
 \bea{ll}
 3 \TEV 
 \lmk \frac{m_I}{10^{12} \GeV} \rmk
 \lmk \frac{\Br_I}{5} \rmk^{-1} 
 \lmk \frac{\lambda}{10^{-11}} \rmk^{-1},
&\quad \text{ for }  n=4, \\
 0.7 \TEV 
 \lmk \frac{m_I}{10^{13} \GeV} \rmk
 \lmk \frac{\Br_I}{5} \rmk^{-1} 
 \lmk \frac{\lambda}{10^{-4}} \rmk^{-1/2}
 \lmk \frac{H_{\text{osc}}}{5 \TEV} \rmk^{-1/2},  &\quad \text{ for }  n=6 \\
  0.4 \TEV 
 \lmk \frac{m_I}{10^{11} \GeV} \rmk
 \lmk \frac{\Br_I}{5} \rmk^{-1} 
 \lmk \frac{\lambda}{10^{-4}} \rmk^{-1/3}
 \lmk \frac{H_{\text{osc}}}{5 \TEV} \rmk^{-2/3},   &\quad \text{ for }  n=8, 
 \eea
 \right.
 \label{grav1 result1}
\eeq
or 
\beq
 \mDM \simeq 
 \left\{
 \bea{ll}
 0.4 \TEV 
 \lmk \frac{\lambda}{10^{-11}} \rmk^{3/2},
&\quad \text{ for }  n=4 ,\\
 3 \TEV 
 \lmk \frac{\lambda}{10^{-6}} \rmk^{3/4}
 \lmk \frac{H_{\text{osc}}}{5 \TEV} \rmk^{3/4},  &\quad \text{ for } n = 6,  \\
 0.2 \TEV 
 \lmk \frac{\lambda}{10^{-4}} \rmk^{1/2}
 \lmk \frac{H_{\text{osc}}}{5 \TEV} \rmk,   &\quad \text{ for }  n = 8, 
 \eea
 \right.
 \label{grav1 result2}
\eeq
where we assume $\epsilon = 1$. 
Eq.~(\ref{grav1 result1}) is the case where the DM abundance is determined by 
the contribution from the decay of inflaton, while
Eq.~(\ref{grav1 result2}) is the case where the DM abundance is determined by 
the contribution from inelastic scattering process during reheating.%
\footnote{
As we have mentioned, the prediction given in Eq.~(\ref{grav1 result2}) involves 
an uncertainty within about one order of magnitude. 
}
Note that $H_{\text{osc}}$ is roughly given by the squark mass 
at the energy scale $\phi_{\text{osc}} \sim 10^{15} \GEV$, 
and the parameter $\lambda$ has an upper bound as $\lambda \lesssim 10^{-4}$ 
to avoid a sizable baryonic isocurvature perturbation as explained in Sec.~\ref{ADBG}.

Interestingly, by combining Eqs.~(\ref{rhoDM dir}) and (\ref{main result}), 
we obtain a lower bound on the inflaton mass as a function of reheating temperature as 
\beq
 m_I \gtrsim 1.3 \times 10^{12} \GEV \lmk \frac{\Br_I}{5} \rmk 
  \lmk \frac{\TR}{0.1 \GEV} \rmk^{5/2}, 
\eeq
to obtain the correct DM density, Eq.~(\ref{rhoDM}). 
Although we can evade this constraint for $\TR \gtrsim m_{\rm DM}$, 
only $n=4$ flat direction with $\lambda \gg 10^{-10}$ is consistent with 
such a high reheating temperature. 
Otherwise, the DM abundance exceeds the observed one 
whatever the DM mass is.

Finally, we comment on the testability of this scenario. 
For SUSY particles with masses of ${\cal O}(1) \TEV$, the elastic 
scattering cross section between the bino-like LSP and nucleon is 
as large as $10^{-46} - 10^{-45}$ cm$^2$, 
which is detectable in future direct detection experiments of DM 
such as XENON1T~\cite{Aprile:2012zx}. 
Since the thermal relic of 
the bino-like LSP is usually much
larger than 
the observed DM abundance, 
the above scenario would be favored if DM is detected in that region.

\subsection{\label{grav with Q}Gravity mediation with Q-balls}

As we have mentioned in the previous subsection, 
the mass of the AD field $m_\phi (\abs{\phi})$ logarithmically depends on $\abs{\phi}$ 
due to the renormalization group running of squark masses. 
We basically write the potential of the AD field 
in gravity mediation as
\begin{equation}
 V = m_\phi^2 (\abs{\phi}) \abs{\phi}^2 
 = m^2_\phi \vert \phi \vert^2 \left( 1+ K \log \frac{ \vert \phi \vert^2}{\Mpl^2} \right),
\end{equation}
where the second term in the parenthesis represents renormalization group running. 
In many cases in gravity mediation, 
the strong interaction dominates quantum corrections for a typical flat direction, 
and we obtain 
$K < 0$ and $\abs{K} \sim 0.01-0.1$~\cite{EnMc}, 
which satisfies the condition for Q-ball formation in Eq.~(\ref{Q condition}). 
The configuration of the AD field is obtained by solving Eq.~(\ref{Q-ball eq}) with the above potential. 
The solution is well approximated by~\cite{EnMc}
\beq
 \phi(r, t ) \simeq \frac{1}{\sqrt{2}} \phi_0 e^{- r^2 / 2 R^2} e^{-i \omega_0 t},
\eeq
where $R$, $\omega_0$, and $\phi_0$ are given as
\begin{eqnarray}
&&R \simeq \frac{1}{ \vert K \vert^{1/2} m_\phi (\phi_0)}, \label{gravproperty} \\
&& \omega_0 \simeq m_\phi (\phi_0), \\
&&\phi_0 \simeq \lmk \frac{\vert K \vert}{\pi} \rmk^{3/4} m_\phi (\phi_0) Q^{1/2},  
\end{eqnarray}
where $m_\phi (\phi_0)$ is the mass of the AD field at the energy scale of $\phi_0$. 
Since the energy of the Q-ball, $M_Q$, is calculated from Eq.~(\ref{E}) as
\beq
 M_Q \simeq m_\phi (\phi_0) Q, 
\eeq
we find that 
the energy of the Q-ball per unit charge is approximately equal to 
$\omega_0 \simeq m_\phi \gg 1 \GEV$ and thus Q-balls decay into quarks.

Using $\omega_0 \simeq m_\phi$ and $R \sim m_\phi^{-1}$ in Eq.~(\ref{Q charge}),
we estimate
a typical charge of Q-ball formed after the Affleck-Dine baryogenesis as 
\beq
  Q  &\sim& \beta \left( \frac{\abs{\phi_{\text{osc}}}}{m_\phi} \right)^{2}, \label{Q in grav 0}\\
  &\simeq& 2 \times 10^{23}  \left( \frac{\abs{\phi_{\text{osc}}}}{3 \times 10^{15} \GeV} \right)^{2}  
  \left( \frac{m_\phi}{1\text{ TeV}} \right)^{-2}. 
  \label{Q in grav}
\eeq
The numerical simulations have shown that 
the coefficient $\beta$ is approximately given by $2 \times 10^{-2}$~\cite{KK1, KK2, Hiramatsu:2010dx},
which we have used in the second line.

The decay rate of the Q-ball is calculated from Eq.~(\ref{Q decay}) 
with $\tilde{R} \simeq R ( 2 \log ( \phi_0 / \sqrt{2} \omega_0 ) )^{1/2} \simeq 7 R$. 
For $n=4$, the Q-ball decays after the reheating as 
\begin{eqnarray}
 \Td 
 \simeq
 10^3 \GEV 
 \lmk \frac{\lambda}{10^{-4}} \rmk^{1/2}
 \lmk \frac{H_{\text{osc}}}{1 \TEV} \rmk^{-1/2}
 \lmk \frac{m_\phi}{1 \TEV} \rmk^{3/2}, \qquad \text{ for } n = 4,
\end{eqnarray} 
where we assume that 
the effective number of relativistic degrees of freedom at the decay time $g_*$ is about $200$, 
and $n_q \sim 10$. 
Using Eq.~(\ref{non dominant cond.}), 
we checked that 
the energy density of Q-balls never dominates the Universe. 
For $n=6$ and $n=8$, the Q-ball decays just before the reheating as 
\beq
 \Td 
 \simeq
 \left\{
 \bea{ll}
 2 \GEV 
 \lmk \frac{\lambda}{10^{-4}} \rmk^{3/8}
 \lmk \frac{H_{\text{osc}}}{1 \TEV} \rmk^{1/8}
 \lmk \frac{m_\phi}{1 \TEV} \rmk^{3/4} , &\qquad \text{ for } n = 6, \\
 60 \MEV 
 \lmk \frac{\lambda}{10^{-4}} \rmk^{1/4}
 \lmk \frac{H_{\text{osc}}}{1 \TEV} \rmk^{1/4}
 \lmk \frac{m_\phi}{1 \TEV} \rmk^{3/4} , &\qquad \text{ for } n = 8,  
 \eea
 \right.
 \label{Q decay grav}
\eeq
where we assume that 
the effective number of relativistic degrees of freedom at the decay time $g_*$ is $10.75$. 
We find that Q-balls decay 
after DM freezes out for $n=6$ and $n=8$.

Let us consider DM production mechanisms.
In order to suppress the baryonic isocurvature perturbation (see Eq.~(\ref{TR})), 
the reheating temperature is required as $\TR \lesssim 1 \GEV$ for 
$n=4$ flat direction with $\lambda \lesssim 10^{-10}$ or 
$n=6$, $8$ flat directions. 
In such a low reheating temperature, 
the two contributions, from the direct decay of the inflaton and from inelastic scatterings
during the thermalization,
explained in the previous subsection 
can generate DM, 
and the results is again given by Eqs.~(\ref{rhoDM dir}) and (\ref{main result}). 
In addition, 
the decay of Q-balls also generate a sizable DM abundance~\cite{EnMc, 
Fujii:2001xp, Fujii:2002kr, Fujii:2002aj, Roszkowski:2006kw, Seto:2007ym, KKY} 
because they decay after DM freezes out (see Eq.~(\ref{Q decay grav})). 

In the rest of this subsection,
we focus on a scenario of baryon and DM co-genesis from Q-ball decay.
This scenario can be realized in the non-shaded region in Fig.~\ref{fig1}
where the two contributions of DM abundance,
from the direct decay of inflaton and from inelastic scatterings,
are negligible.
This is the case with, for example, a flat direction with $n=8$ and $\lambda \lesssim 10^{-4}$.

As we have explained in Sec.~\ref{Q-ball}, 
the decay rate of Q-ball is saturated and determined by the Pauli blocking effect. 
While SUSY particles are produced 
from the Q-ball surface 
only through elementary decay process like $\tilde{q} \to q + (\text{gaugino})$, 
quarks are dominantly produced through scattering process via gaugino or Higgsino exchange 
like $\tilde{q} + \tilde{q} \to q + q$~\cite{KY}. 
Thus, the ratio 
of the Q-ball decay into
sparticles and quarks is calculated as 
\beq
 \frac{\Br(\text{Q-ball} \to (\text{gauginos)})}{\Br(\text{Q-ball} \to q)} \simeq \frac{n_s}{8 n_q},
\eeq
where a factor of $8$ is due to
the difference of the elementary processes as we have explained in Sec.~\ref{Q-ball}. 
The factor $n_s$ is the effective number of sparticles into which Q-balls can decay. 
Since the flux of massive particles from a Q-ball surface 
is smaller than that of massless particles, 
there is a correction
due to non-zero sparticle masses, 
which is derived in the Appendix. 
Thus we obtain
\beq
 n_s = \sum_s g_s f (m_{\text{s}} / \omega_0), 
 \label{n_g}
\eeq
where $m_s$ is the mass of the sparticle $s$,
$g_s$ is the number of species for the sparticle
and $f$ is a function given in the Appendix. 
For example, $g_s = 1$, $3$, and $8$ for the bino, wino, and gluino, respectively.%
\footnote{
Note that $g_s$ also depends on the flat direction.
In the case of $\bar{u} \bar{d} \bar{d}$ flat direction, 
Q-balls do not decay into winos, $g_\text{wino} = 0$,
since they
consists of only right-handed squarks~\cite{KKY}. 
}
However, Q-balls can decay only into particles lighter than 
the energy of the Q-ball per unit charge, $\omega_0$, which is approximately equal to 
the mass of squarks at the energy scale of $\abs{\phi}$ ($\sim 10^{15} \GEV$). 
In a typical model of gravity mediation, 
a mass of squarks at the energy scale of $10^{15} \GEV$ 
is mostly smaller than the mass of the gluino
and larger than that of the bino (LSP)~\cite{KKY in prep}.
Thus in the typical models 
Q-balls can decay into binos, 
and not into gluino.%
\footnote{
Even in this case Eq.~(\ref{n_g}) is valid since $f(x > 1) = 0$.
}
Depending on 
a model, 
Q-balls can also decay into winos and Higgsinos. 
Hereafter, we assume Q-balls can not decay into Higgsinos, 
for simplicity (see Ref.~\cite{KKY in prep} for complete discussion).

Since all sparticles eventually decay into binos,%
\footnote{
For the case of the axino LSP, see Refs.~\cite{Roszkowski:2006kw, Seto:2007ym}.
}
we obtain the following formula for the baryon-to-DM ratio:
\beq
 \frac{\Omega_{\text{DM}}}{\Omega_b} &=& \frac{m_{\tilde{b}}}{\abs{b} m_p} \frac{n_s}{8 n_q}, \\
 &\simeq& \frac{m_{\tilde{b}}}{b m_p} \frac{ \sum_s f (m_{\text{s}} / \omega_0) 
}{8 n_q}. 
 \label{bDMratio}
\eeq
We should emphasize that this ratio is independent of the reheating temperature and 
the charge of Q-balls. 
As an illustration, 
let us calculate two asymptotic solutions.
When $m_{s} \ll \omega_0$, 
the function $f$ approaches $1$ (see the Appendix) and $n_s \simeq 12$, where we have included contributions from gauginos. 
On the other hand, if 
$m_{\tilde{b}} \to \omega_0$ with other particles mass larger than $\omega_0$,
a combination of $m_{\tilde{b}} n_s / \omega_0$ approaches 
$4 (1 - m_{\tilde{b}} / \omega_0)^3 \ll 1$. 
Thus we obtain two asymptotic solutions for the bino mass, which yields the correct ratio of DM and baryon density;
\beq
 m_{\tilde{b}} \approx
 \left\{
 \bea{ll}
  0.7 n_q \abs{b} m_p \frac{\Omega_{\text{DM}}}{\Omega_b} \simeq 10 \GEV \vspace{0.3cm}~~(\omega_0  \gg m_s)\\ 
  \omega_0 \lkk  1 -  \lmk 2 n_q \abs{b} \frac{\Omega_{\text{DM}}}{\Omega_b}  
  \frac{m_p}{ \omega_0} \rmk^{1/3} \rkk  ~~(m_{\tilde{b}} \to \omega_0),
 \eea
 \right.
\eeq
where we assume $n_q =10$ and $b = 1/3$ in the first line. 
The bino mass of $10 \GEV$ is unrealistic for ordinary models in gravity mediation. 
We conclude that 
we can explain the observed baryon-to-DM ratio
if the bino mass is close to below $\omega_0$.

\begin{figure}[t]
\centering 
\includegraphics[width=.65\textwidth, bb=0 0 360 249]{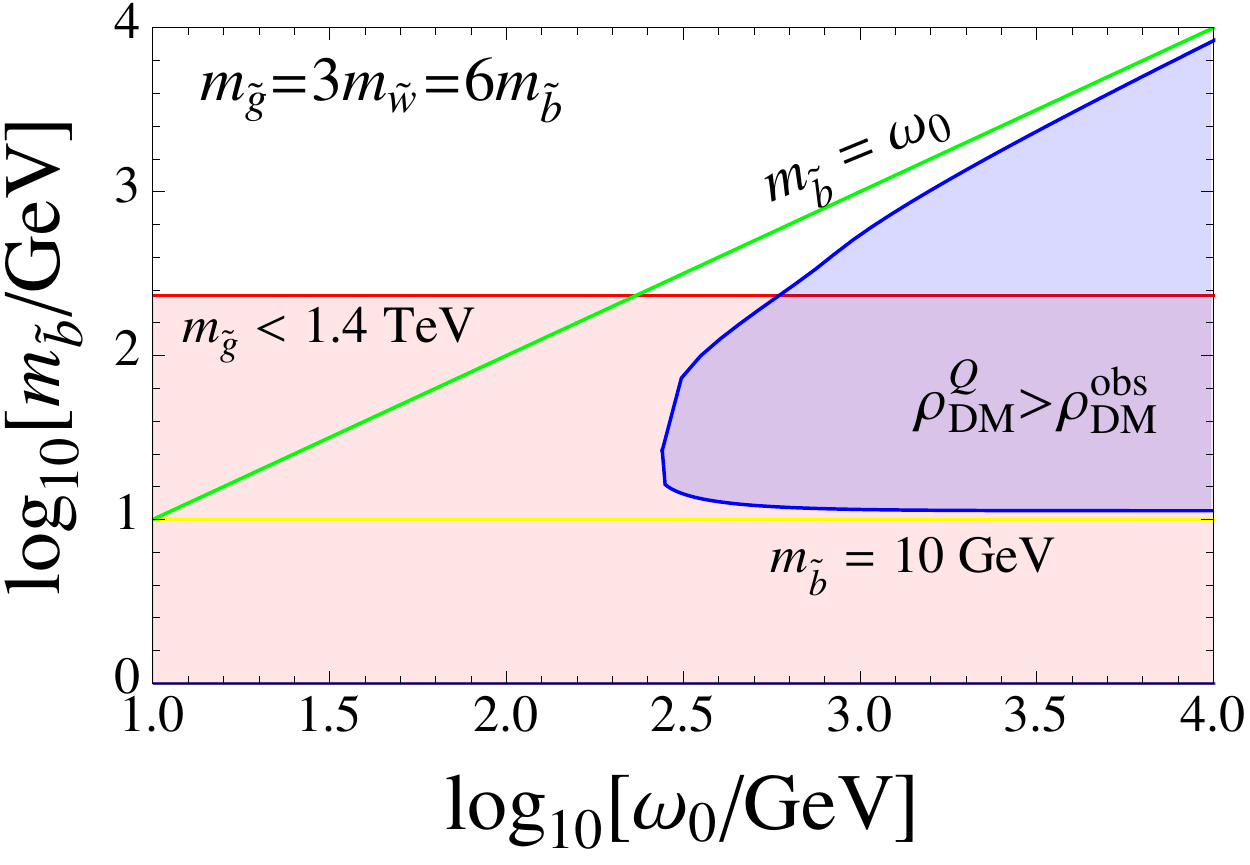} 
\caption{
Exclusion plot in a model of gravity mediation with Q-ball formation. 
We assume that 
the number of quarks interacting with Q-ball, $n_q$, is $10$ and 
baryon charge of the AD field, $b$, is $-1/3$. 
We also assume that the masses of the wino and gluino 
are two and six times larger than that of the bino, which is a typical case 
in gravity mediation with the grand unified theory relation. 
The abundance of DM produced from
Q-ball decay ($\rho_{\text{DM}}^Q$) 
is larger than that observed in the blue (dark gray) shaded region. 
In the red (light gray) shaded region, the gluino mass is smaller than $1.4$ TeV 
and excluded by the gluino search at the ATLAS Collaboration~\cite{gluino}.
The green-dotted and yellow-dashed lines indicate
the limit of $m_{\tilde{b}}\to \omega_0$ and $m_{\tilde{b}}\to 10$ GeV, respectively.
}
  \label{fig2}
\end{figure}

In Fig.~\ref{fig2}, we show the constraint on $\omega_0$ and $m_{\tilde{b}}$.
On the boundary of and inside the blue (dark gray) shaded region, 
the DM density produced by the decay of Q-balls 
is equal and larger than the observed value respectively.
Here, we have assumed the grand unified theory (GUT) relation, where the masses of the wino and gluino 
are two and six times larger than that of the bino.
Under the GUT relation, 
the red (light gray) shaded region is already excluded 
by the gluino search at the ATLAS Collaboration~\cite{gluino}.
Detailed calculations in the constrained MSSM 
will be performed in the near future~\cite{KKY in prep}.

\subsection{\label{GMSB without Q}Gauge mediation without Q-ball}

In models of gauge mediation, 
SUSY breaking in a hidden sector is transmitted to the standard model sector 
by gauge interactions 
mediated by a messenger sector. 
Since the AD field has some gauge charge, 
gauge fields acquire effective masses of the order of $g \abs{\phi}$, 
where $g$ generically stands for the Standard Model gauge coupling. 
The transmission of SUSY breaking effect is therefore suppressed for 
$g \abs{\phi} \gg M_s$,
where $M_s$ is a messenger scale~\cite{GMM}, and thereby
the soft mass of the AD field is suppressed.
The condition for Q-ball formation, Eq.~(\ref{Q condition}), is therefore satisfied
if there is no term in the potential of the AD field 
other than the soft mass term. 
However, 
if the AD field is $L H_u$ flat direction, 
the $\mu$ term of Higgs multiplets 
prevents the flat direction forming Q-balls~\cite{Fujii:2001dn}.%
\footnote{
In the model of seesaw mechanism, $F$ term potential of right-handed neutrinos also lift $LH_u$ flat direction.
Since an upper bound on isocurvature fluctuation requires 
large value of the AD field, the F term potential should be sufficiently small.
The smallness is realized if right-handed neutrinos are light and/or one of left-handed neutrinos is light.
}
In this subsection, we consider the case of $L H_u$ flat direction in gauge mediation
and construct a consistent scenario to account for the observed DM density as well as 
baryon density.

Since the mass of the gravitino $m_{3/2}$ is much smaller than the electroweak scale, 
and since the AD field starts to oscillate at $H_{\text{osc}} \sim \mu \sim 1$ TeV, 
the ellipticity parameter defined in Eq.~(\ref{ellipticity}) is much smaller than one, 
\beq
 \epsilon \simeq 5 \times 10^{-3} \lmk \frac{\mg}{5 \GEV} \rmk 
 \lmk \frac{\mu}{1 \TEV} \rmk^{-1}. 
\eeq
Taking into account 
$\Delta B_\text{after sphaleron} \simeq 0.3 \times  \Delta (B-L)_\text{before sphaleron}$ 
due to the sphaleron effect~\cite{Fukugita:1986hr}
and $b = -1/2$ for $B-L$ charge of $L H_u$ flat direction, 
we find 
that Eq.~(\ref{TR})
is 
rewritten as 
\beq
\TR &\simeq& 
\left\{
\bea{ll}
1\times 10^8 \GEV 
\lmk \frac{\mg}{6 \GEV} \rmk^{-1}
\lmk \frac{\mu}{1 \TEV} \rmk 
\lmk \frac{\lambda}{10^{-4}} \rmk &\qquad \text{ for } n=4, \\
2 \times 10^5 \GEV 
\lmk \frac{\mg}{8 \MEV} \rmk^{-1}
\lmk \frac{\mu}{1 \TEV} \rmk^{3/2} 
\lmk \frac{\lambda}{10^{-4}} \rmk^{1/2} &\qquad \text{ for } n=6, \\
2 \times 10^4 \GEV 
\lmk \frac{\mg}{0.9 \MEV} \rmk^{-1}
\lmk \frac{\mu}{1 \TEV} \rmk^{5/3}
\lmk \frac{\lambda}{10^{-4}} \rmk^{1/3} &\qquad \text{ for } n=8. 
\eea
\right.
\label{TR gauge med}
\eeq
Note that we assume a discrete R-symmerty 
which controls higher-dimensional terms in the superpotential as well as the term with $n=2$.
Therefore, in contrast to the case with the R-patiry, $LH_u$ flat direction with $n=6$ or $8$ is also possible.

Next, let us estimate the abundance of the gravitino, which is the LSP in gauge mediation.
There are two contributions to the gravitino production, depending on the reheating temperature.
If the reheating temperature is larger than the mass scale of MSSM particles, MSSM particles are in the thermal equilibrium and 
gravitinos are produced from thermal bath through scatterings between 
gluons and gluinos, 
as is discussed in Refs.~\cite{Moroi:1993mb, Bolz:2000fu, 
Pradler:2006qh, Kusenko:2010ik, Mambrini:2013iaa}.
Here we quote the result from Ref.~\cite{Bolz:2000fu}: 
\beq
 \frac{\rho_{3/2}^{\text{th}}}{s} 
 \simeq 
 0.4 \EV 
 \lmk \frac{\TR}{10^8 \GEV} \rmk
 \lmk \frac{m_{\tilde{g}} }{1 \TEV } \rmk^{2}
 \lmk \frac{\mg}{2 \GEV } \rmk^{-1}. 
 \label{rho32 thermal}
\eeq
where 
$m_{\tilde{g}}$ is the gluino mass. 
The parameter dependences of this result can be understood by 
\beq
 \frac{\rho_{3/2}^{\text{th}}}{s} 
&\sim& \left. 
 \frac{\mg \la \sigma_{\rm scatt} v \ra n_r^2}{s H}
 \right\vert_{T = \TR}, \\
  \la \sigma_{\rm scatt} v \ra &\sim& \frac{g_3^2 m_{\tilde{g}}^2}{\mg^2 \Mpl^2}, 
\eeq
where $n_r = \zeta(3) T^3 /\pi^2$ is the number density of radiation in the thermal bath, 
and $g_3$ is a strong coupling constant.

If the reheating temperature is smaller than the mass scale of MSSM particles, MSSM particles are produced by the mechanism explained in Sec.~\ref{grav without Q}. 
Then, MSSM particles eventually decay into gravitinos.
The abundance of the gravitino is given by
\beq
 \frac{\rho_{3/2}^{\rm inela}}{s} &=&
 \mg \left(\frac{T_{\rm RH}}{m_{\rm SUSY}}\right)^3
, \\
 &\simeq& 0.3 \EV 
 \lmk \frac{\TR}{5 \GEV} \rmk^3
 \lmk \frac{m_{\rm SUSY} }{5 \TEV } \rmk^{-3}
 \lmk \frac{\mg}{300\MEV} \rmk,
 \label{rho32 inela}
\eeq
where $m_{\rm SUSY}$ is the mass scale of MSSM particles whose production by inelastic process is efficient.
Typically, $m_{\rm SUSY}$ can be identified with the mass of the NLSP.
Note that if the NLSP is produced too much, 
its annihilation is effective and the gravitino abundance decreases accordingly. 
When the NLSP is 
right-handed stau,
which is the case with typical gauge mediation models,
its annihilation cross section is given as~\cite{Feng:2004mt} 
\beq
 \la \sigma_{\text{ann}} v \ra_{\tilde{l}_R} 
 \simeq
 3 \times 10^{-11} \GEV^{-2} \lmk \frac{m_{\tilde{l}_R}}{5 \TEV}\rmk^{-2}, 
 \label{NLSP sigmav}
\eeq
where 
we have omitted bino mass dependence 
for simplicity. 
In analogy with the derivation of Eq.~(\ref{rho_anni}), 
the annihilation effect of the NLSP puts 
an upper bound on the gravitino abundance produced from the decay of the NLSP: 
\beq
 \frac{\rho_{\text{DM}}^{\text{inela (ann)}}}{s} 
 &\simeq& 
 \sqrt{\frac{45}{8 \pi^2 g_*}} \frac{\mg}{\TR \la \sigma_{ann} v \ra \Mpl}, \nonumber\\
 &\simeq& 0.3 \EV 
 \lmk \frac{\TR}{10 \GEV} \rmk^{-1}
 \lmk \frac{\mg}{1 \GEV} \rmk
  \lmk \frac{m_{\tilde{l}_R}}{5 \TEV} \rmk^2. 
 \label{rho_anni_GM}
\eeq
In summary, the gravitino abundance is given by
\beq
 \frac{\rho_{3/2}}{s} 
 \simeq 
 \left\{
 \bea{ll}
 \displaystyle{
  \frac{\rho_{3/2}^{\text{th}}}{s} 
  }
 &\qquad \text{ for } \TR \gtrsim m_{\rm SUSY}, \\
 \min \lkk 
\displaystyle{
 \frac{\rho_{\text{DM}}^{\text{inela}}}{s}, 
  \frac{\rho_{\text{DM}}^{\text{inela (ann)}}}{s} 
  }
 \rkk
 &\qquad \text{ for } \TR \ll m_{\rm SUSY}.
  \eea
 \right.
 \label{gauge without Q result}
\eeq

Note that gravitinos and MSSM particles can be produced in inflaton decay 
as is discussed in Sec.~\ref{grav without Q}. 
If the reheating temperature is lower than the mass scale of MSSM particles,
the gravitino abundance is given by Eq.~(\ref{rhoDM dir}). 
We can neglect this contribution at least if $m_\phi \gtrsim 10^{12} \GEV$. 
On the other hand, if the reheating temperature is higher than the mass scale of MSSM particles, 
the MSSM particles are soon thermalized and Eq.~(\ref{rho32 thermal}) holds.
The gravitino abundance produced in inflaton decay depends on 
the branching ratio of inflaton decay into gravitino. 
In this subsection, we assume that 
the branching into gravitino is so small that 
its abundance from inflaton decay is much less than the observed DM abundance.

\begin{figure}[t]
\centering 
\includegraphics[width=.45\textwidth, bb=0 0 360 365]{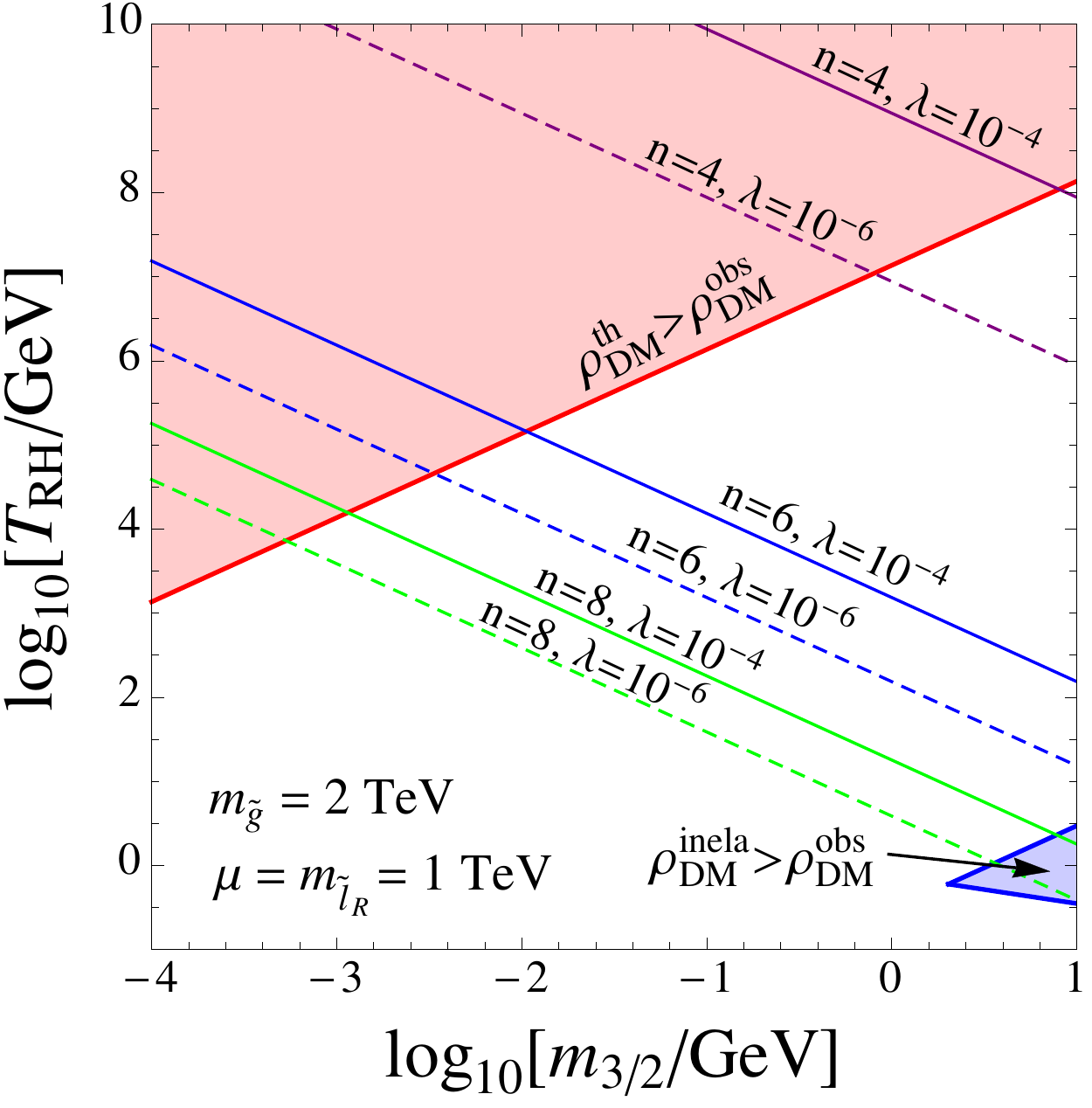} 
\caption{
Exclusion plot in a model of gauge mediation without Q-ball formation. 
We assume that 
the mass of gluino is $2 \TEV$, 
$\mu = 1 \TEV$, 
and $m_{\rm SUSY} = m_{\rm NLSP} = 1 \TEV$. 
We also assume that the NLSP is right-handed stau and 
its annihilation cross section is given by 
Eq.~(\ref{NLSP sigmav}). 
The abundance of DM produced from 
scatterings between gluon and gluinos in the thermal bath ($\rho^{\text{th}}_{\text{DM}}$) 
is larger than that observed in the red (light gray) shaded region.
In the blue (dark gray) shaded region, 
the abundance of DM produced by the decay of the NLSP 
produced through inelastic scatterings during the thermalization process exceeds the observed one.
The (dashed) lines 
are the reheating temperature required by the success of the Affleck-Dine baryogenesis 
for $n=4$, $6$, and $n=8$ flat direction with $\lambda = 10^{-4}$ ($10^{-6}$). 
}
  \label{fig3}
\end{figure}

In Fig.~\ref{fig3}, we show the constraint on the gravitino mass and the reheating temperature.
We assume that 
the mass of gluino is $2 \TEV$, 
$\mu = 1 \TEV$,
and $m_{\rm SUSY} = m_{\rm NLSP} =  1 \TEV$. 
We also assume that 
the NLSP is right-handed stau and its annihilation cross section is given by 
Eq.~(\ref{NLSP sigmav}). 
On the boundary of and inside the shaded region, the DM abundance is 
equal to and larger than that observed, respectively.
In the red (light gray) shaded region, DM is produced from scatterings betweens gluons and gluinos. 
The upper bound on the blue (dark gray) shaded region is determined by the NLSP 
abundance produced through inelastic scatterings 
(that is, Eq.~(\ref{rho32 inela}) with $\TR$ given by Eq.~(\ref{TR gauge med})), 
while the lower bound on that is determined by the annihilation of the NLSP (Eq.~(\ref{rho_anni_GM})). 
The (dashed) lines show the required reheating temperature and the gravitino mass for the successful Affleck-Dine Baryogenesis, for $n=4$, $6$, and $8$ with $\lambda = 10^{-4}$  ($10^{-6}$).

If the DM abundance is produced by scatterings from thermal bath 
(the lower boundary of the red (light gray) shaded region in Fig.~\ref{fig3}), from Eqs.~(\ref{TR gauge med}) and (\ref{rho32 thermal}),
the mass of the gravitino is predicted to be
\beq
 \mg \simeq 
 \lmk \frac{\mu }{ 1 \TEV} \rmk
 \lmk \frac{ m_{\tilde{g}}}{2 \TEV} \rmk \times 
 \left\{
 \bea{ll}
 8 \GEV \lmk \frac{\lambda}{10^{-4}} \rmk^{1/2} &\qquad \text{ for } n=4. \\
 10 \MEV \lmk \frac{\lambda}{10^{-4}} \rmk^{1/4} &\qquad \text{ for } n=6, \\
 1 \MEV \lmk \frac{\lambda}{10^{-4}} \rmk^{1/7} &\qquad \text{ for } n=8.
 \eea
 \right.
\eeq
The case with $n=4$ 
and 
$\lambda \sim 10^{-4}$
requires large gravitino mass. 
The large gravitino mass contributes to electric dipole moments (EDMs)
through supergravity effect in general. 
For gravitino mass larger than ${\cal O}(100) \MEV$, 
suppression of 
CP phases due to the supergravity effect is necessary~\cite{Moroi:2011fi}.

Let us consider the case where the DM abundance is explained by the contribution from inelastic scatterings during the thermalization process 
(the boundary of the blue (dark gray) shaded region in Fig.~\ref{fig3}), focusing 
on the free-streaming velocity of the gravitino.
The NLSP decays into gravitinos
with a decay rate given by
\beq
 \Gamma_{\text{NLSP}} \simeq \frac{\mN^5}{48 \pi \mg^2 \Mpl^2}. 
\eeq
The decay temperature of the NLSP is thus given by
\beq
 T_{\text{NLSP}} \simeq 
 0.3 \GEV 
 \lmk \frac{\mN}{5 \TEV} \rmk^{5/2}
 \lmk \frac{\mg}{0.2 \GEV} \rmk^{-1}. 
 \label{NLSP decay temp}
\eeq
The present-day free-streaming velocity of the gravitino DM
is calculated as~\cite{ShKu, Jedamzik:2005sx, Steffen:2006hw}
\beq
 v_0 &\simeq& \frac{ \mN/2 }{\mg} \frac{T_0}{T_{\text{NLSP}}} 
 \lmk \frac{g_{*s} (T_0)}{g_{*s} ( T_{\text{NLSP}} )} \rmk^{1/3}, \\
 &\simeq& 
  9 \times 10^{-9} \lmk 
  \frac{\mN}{5 \text{ TeV}} \rmk^{-3/2}
 \lmk \frac{g_{*s} (T_0)}{g_{*s} ( T_{\text{NLSP}} )} \rmk^{1/3}, 
 \label{free streaming}
\eeq
where $T_0$ ($\simeq 2.3 \times 10^{-4}\EV$) is the temperature at the present time,
and $g_{*s}$ is the effective number of relativistic degrees of freedom for entropy.
The observation of the Lyman-$\alpha$ forest puts an upper bound on the free-streaming velocity
as $v_0 \lesssim 2.5 \times 10^{-8}$~\cite{Viel:2013fqw} (see Ref.~\cite{Steffen:2006hw} for review). 
This results in 
$\mN \gtrsim 3 \TEV$
from Eq.~(\ref{free streaming}). 
It is expected that future observations of redshifted 21 cm line would improve the upper bound
to $v_0 \lesssim 2 \times 10^{-9}$~\cite{Sitwell:2013fpa}. 
Hence this scenario
would be tested by future observations of redshifted 21 cm line 
if the mass of the NLSP is less than about $14 \TEV$. 

Finally, we comment on the BBN bound on the gravitino mass~\cite{Kawasaki:2004qu, Kawasaki:2008qe}. 
Since the NLSP is long-lived in gauge mediation, 
they may produce high energy hadrons or photons after the BBN epoch. 
These high energy particles induce hadro- and photo-dissociation processes of light elements, 
which spoil the success of the BBN scenario. 
This puts a stringent lower bound on the NLSP mass especially for the case of $\mg \gtrsim 1 \GEV$. 
When the NLSP is the stau, which is a typical case in gauge mediation, 
the NLSP mass has to be larger than 
about $700 \GEV$ for $\mg \simeq 10 \GEV$~\cite{Kawasaki:2008qe}.

\subsection{\label{GMSB with Q}Gauge mediation with Q-balls}

In this subsection, we consider a flat direction other than $L H_u$ flat direction. 
As explained in the previous subsection, 
the soft mass term for the AD field is absent for a VEV larger than the messenger scale 
because
the transmission of SUSY breaking effect is suppressed for 
such a large VEV. 
Thus we obtain the following potential for the AD field~\cite{GMM}: 
\begin{equation}
V = 
\left\{
\bea{ll}
 m^2_\phi \abs{\phi}^2  &\qquad \text{ for }  g \abs{\phi} \ll M_s, \\
 M_F^4  \left[ \log \frac{g^2 \vert \phi \vert^2 }{M_s^2}  \right]^2 &\qquad \text{ for } g \abs{\phi} \gg M_s, 
\eea
\right.
\label{GMSB V}
\end{equation}
where $M_s$ 
is a messenger mass, 
and $g$ generically stands for the Standard-Model gauge coupling. 
A parameter $M_F^2$ is proportional to the VEV of the $F$ component of 
a gauge-singlet chiral multiplet in the messenger sector as 
\beq
 M_F^2 = \frac{m_\phi^2 M_s^2}{g^2} = \frac{g y}{(4 \pi)^2} \la F_s  \ra,
  \label{M_F}
\eeq
where $y$ is a coupling constant for the interaction between 
the gauge singlet chiral multiplet and the messenger field. 
The mass of the gravitino is related to the SUSY breaking $F$-term as
\beq
 \la F_s \ra &=& k \sqrt{3} \mg \Mpl,  \label{F_s} \\
 k &\le& 1, 
\eeq
where a factor $k$ is less than one 
when 
the messenger sector 
indirectly couple to the SUSY breaking sector. 
Hereafter we redefine the combination of $yk$ as $k$ ($\lesssim 1$).

Since the curvature of the potential at the energy scale of 
$\abs{\phi_{\text{osc}}}$ is roughly given by $M_F^2 / \abs{\phi_{\text{osc}}}$, 
the AD filed begins to oscillate at the time of $H_{\text{osc}} \sim M_F^2 / \abs{\phi_{\text{osc}}}$ ($\sim \omega_0$). 
Using this and Eqs.~(\ref{VEV osc}), (\ref{M_F}), and (\ref{F_s}), the VEV of the AD field at the beginning of its oscillation is calculated as 
\beq
\abs{\phi_{\text{osc}}} 
\simeq 
\left\{
\bea{ll}
3 \times 10^{13}
\ k^{1/3}
\lmk \frac{\mg}{1 \GEV} \rmk^{1/3}
\lmk \frac{\lambda}{10^{-6}} \rmk^{-1/3}, 
&\qquad \text{ for } n = 4, \\
10^{15}
\ k^{1/5}
\lmk \frac{\mg}{1 \GEV} \rmk^{1/5}
\lmk \frac{\lambda}{10^{-4}} \rmk^{-1/5}, &\qquad \text{ for } n = 6, \\
10^{16}
\ k^{1/7}
\lmk \frac{\mg}{1 \GEV} \rmk^{1/7}
\lmk \frac{\lambda}{10^{-4}} \rmk^{-1/7}, &\qquad \text{ for } n = 8.
\eea
\right.
\label{phi_osc gauge}
\eeq
We can calculate 
the ellipticity parameter $\epsilon$ ($\simeq \mg/H_{\text{osc}}$) and 
the required reheating temperature as 
\beq
\epsilon \simeq
\left\{
\bea{ll}
10^{-3} 
\ k^{-2/3}
\lmk \frac{\mg}{1 \GEV} \rmk^{1/3} 
\lmk \frac{\lambda}{10^{-6}} \rmk^{-1/3}, 
&\qquad \text{ for } n = 4, \\
4 \times 10^{-2} 
\ k^{-4/5}
\lmk \frac{\mg}{1 \GEV} \rmk^{1/5} 
\lmk \frac{\lambda}{10^{-4}} \rmk^{-1/5}, 
&\qquad \text{ for } n = 6, \\
0.4 
\ k^{-6/7}
\lmk \frac{\mg}{1 \GEV} \rmk^{1/7} 
\lmk \frac{\lambda}{10^{-4}} \rmk^{-1/7}, 
&\qquad \text{ for } n = 8, 
\eea
\right.
\label{epsilon GMSB 2}
\eeq
and
\beq
\TR \simeq
\left\{
\bea{ll}
3 \times 10^{6} \GEV
\ k^{2/3}
\lmk \frac{\mg}{1 \GEV} \rmk^{-1/3}
\lmk \frac{\lambda}{10^{-6}} \rmk^{4/3} ,
&\qquad \text{ for } n = 4, \\
3 \GEV
\ k^{6/5}
\lmk \frac{\mg}{1 \GEV} \rmk^{1/5}
\lmk \frac{\lambda}{10^{-4}} \rmk^{4/5},
&\qquad \text{ for } n = 6, \\
0.5 \MEV
\ k^{10/7}
\lmk \frac{\mg}{1 \GEV} \rmk^{3/7}
\lmk \frac{\lambda}{10^{-4}} \rmk^{4/7} ,
&\qquad \text{ for } n = 8, 
\eea
\right.
\label{TR gauge with Q}
\eeq
respectively. 
For $n=8$ flat direction, 
the reheating temperature is so small 
that the success of the BBN is spoiled. 
Hereafter, we consider only the cases with $n=4$ and $6$.

The reheating temperature for $n=4$ flat direction 
may be so large that gravitino is produced from the thermal plasma.
From Eq.~(\ref{rho32 thermal}), 
we obtain the gravitino abundance produced from the thermal plasma as 
\beq
 \frac{\rho_{3/2}^{\text{th}}}{s} 
 \simeq 0.1 \EV 
 \ k^{2/3}
 \lmk \frac{m_{\tilde{g}} }{2 \TEV } \rmk^{2}
 \lmk \frac{ \mg}{1 \GEV } \rmk^{-4/3}
 \lmk \frac{\lambda}{10^{-6}} \rmk^{4/3}
 \qquad \text{ for } n = 4. 
\eeq
Although gravitinos and MSSM particles can be produced from inflaton decay directly 
in the same way in Sec.~\ref{grav without Q}, 
we neglect its contribution assuming a large inflaton mass. 
For $n=6$ flat direction, 
the reheating temperature is smaller than the freeze-out temperature of the NLSP
and thus the NSLPs are produced at the reheating epoch 
as explained in the previous subsection (see Eq.~(\ref{rho32 inela})). 
This is also the case with $n=4$ and $\lambda \ll 10^{-9}$. 
Since the NLSPs may annihilate and 
eventually decay into gravitinos, 
the abundance of the gravitino is again given by the 
second line of Eq.~(\ref{gauge without Q result}) 
with the reheating temperature of Eq.~(\ref{TR gauge with Q}). 
Thus we obtain the abundance of the gravitino as 
\beq
 \frac{\rho_{3/2}^{\rm inela}}{s} 
 \simeq 
\left\{
\bea{ll}
 4 \EV 
 \ k^{2}
 \lmk \frac{m_{\rm NLSP} }{10 \TEV } \rmk^{-3}
 \lmk \frac{\lambda}{10^{-10}} \rmk^{4}
  &\qquad \text{ for } n = 4, \\
 0.2 \EV 
 \ k^{18/5}
 \lmk \frac{m_{\rm NLSP} }{5 \TEV } \rmk^{-3}
 \lmk \frac{\mg}{1 \GEV} \rmk^{8/5}
 \lmk \frac{\lambda}{10^{-4}} \rmk^{12/5}
  &\qquad \text{ for } n = 6,
\eea
\right.
\eeq
with an upper bound due to the annihilation of the NLSP as 
\beq
 \frac{\rho_{3/2}^{\text{inela } (\text{ann})}}{s} 
 \simeq 
 \left\{
\bea{ll}
 0.4 \EV 
 \ k^{-2/3}
 \lmk \frac{\mg}{1 \GEV} \rmk^{4/3}
  \lmk \frac{m_{\text{NLSP}}}{7 \TEV} \rmk^2
   \lmk \frac{\lambda}{10^{-10}} \rmk^{-4/3}
  &\qquad \text{ for } n = 4, \\
 0.4 \EV 
 \ k^{-6/5}
 \lmk \frac{\mg}{1 \GEV} \rmk^{4/5}
  \lmk \frac{m_{\text{NLSP}}}{3 \TEV} \rmk^2
   \lmk \frac{\lambda}{10^{-4}} \rmk^{-4/5}
   &\qquad \text{ for } n = 6. 
   \eea
\right.
 \label{rho_anni_GM 2}
\eeq
In addition, Q-balls may provide another source of gravitinos as shown below.

For the potential in Eq.~(\ref{GMSB V}), 
there exists a Q-ball solution, 
approximated to be~\cite{Dvali:1997qv}%
\footnote{
If $\phi_{\text{osc}}$ or $\phi_0$ is 
less than about the messenger mass $M_s$ ($\simeq g M_F^2 / m_\phi$), 
the suppression on the transmission of SUSY breaking effect
is absent and the situation is 
similar to models of gravity mediation~\cite{Doddato:2011fz, Doddato:2011hx}. 
We have checked that 
$\phi_{\text{osc}}$ and $\phi_0$ is larger than $M_s$ in the case we are interested in, 
if the mass of the AD field $m_\phi$ is larger than $10 \TEV$ or 
that of gravitino $\mg$ is less than $10 \GEV$. 
}%
\footnote{
If $\phi_0 \gtrsim M_F^2 / \mg$, 
the potential of the AD field is dominated by 
the soft mass of the form $\mg^2 |\phi|^2$, 
which is induced by gravity mediated SUSY breaking effect. 
In this case, a Q-ball solution is known as a ``new type Q-ball"~\cite{Kasuya:2000sc}, 
which is stable and is a DM candidate for the case of $ \mg / \abs{b} \lesssim 1 \GEV$. 
In this paper, we focus on the case of the LSP DM 
and leave that case for a future work. 
Note that if $k\sim1$, 
a ``new type Q-ball" is never formed. 
}
\beq
 \phi(r, t) \simeq 
 \left\{
 \bea{ll}
 \frac{1}{\sqrt{2}} \phi_0 \frac{\sin (\omega_0 r)}{\omega_0 r} e^{-i\omega_0t} 
 &\quad \text{ for } r \le R \equiv \frac{\pi}{\omega_0}, \\
 0 &\quad \text{ for } r > R ,
 \eea
 \right. 
\eeq
where $\omega_0$ and $\phi_0$ are given as
\beq
\omega_0 &\simeq& \sqrt{2 } \pi M_F Q^{-1/4},  \label{omega_0 GMSB}\\ 
\phi_0 &\simeq& M_F  Q^{1/4}. 
\eeq
We omit ${\cal O}(1)$ logarithmic corrections on Q-ball parameters for simplicity~\cite{Hisano:2001dr}.
Since the energy of the Q-ball is calculated as
\beq
M_Q &\simeq& \frac{4 \sqrt{2} \pi}{3} M_F Q^{3/4}, 
\eeq
the energy of the Q-ball 
per unit charge $\dd M_Q / \dd Q$ is approximated to be $\omega_0$.

Using $R \simeq \pi / \omega_0$ and $\omega_0 \sim M_F^2 / \phi$ in Eq.~(\ref{Q charge}), 
a typical charge of Q-balls formed after the Affleck-Dine baryogenesis 
is estimated as 
\beq
 Q &\sim& \beta \lmk \frac{\abs{\phi_{\text{osc}}}}{M_F} \rmk^4,  \label{Q in gauge 0}\\
  &\simeq& 
  \left\{
  \bea{ll}
  10^{21} 
  \ k^{-2/3}
  \lmk \frac{\mg}{1 \GEV} \rmk^{-2/3}
 \lmk \frac{\lambda}{10^{-9}} \rmk^{-4/3} 
 &\qquad \text{ for } n = 4, \\
  8 \times 10^{21} 
\ k^{-6/5}
  \lmk \frac{\mg}{10 \GEV} \rmk^{-6/5}
 \lmk \frac{\lambda}{10^{-4}} \rmk^{-4/5} 
 &\qquad \text{ for } n = 6, 
 \eea
 \right.
 \label{Q in gauge}
\eeq
where $\beta$ has been calculated 
as $6 \times 10^{-5}$ for $\epsilon \ll 1$ 
by the numerical simulation of Q-ball formation~\cite{KK3}. 
We have used 
Eq.~(\ref{phi_osc gauge}) 
in the last line.%
\footnote{
Q-balls with baryon charge less than about $10^{18}$ 
is completely disappeared 
by the dissipation effect in the thermal plasma~\cite{Laine:1998rg, Banerjee:2000mb} 
(see also Ref.~\cite{KK3}). 
In this case Q-balls do not affect the DM abundance. 
}
From Eqs.~(\ref{Q decay}) and (\ref{Q in gauge}), 
the Q-ball decay temperature is calculated as
\begin{eqnarray}
  \Td
\simeq
  \left\{
  \bea{ll}
 7 \GEV
 \ k^{2/3}
   \lmk \frac{\mg}{1 \GEV} \rmk^{2/3}
 \lmk \frac{\lambda}{10^{-9}} \rmk^{5/6}, 
  &\qquad \text{ for } n = 4, \\
 3 \GEV
\ k
   \lmk \frac{\mg}{10 \GEV} \rmk
 \lmk \frac{\lambda}{10^{-4}} \rmk^{1/2}, 
  &\qquad \text{ for } n = 6, 
  \eea
  \right.
\label{Q_decay temp}
\end{eqnarray} 
where we use $\tilde{R} \simeq R = \pi/\omega_0$ 
and also use Eqs.~(\ref{M_F}), (\ref{F_s}), and (\ref{omega_0 GMSB}) 
to eliminate $\omega_0$ and $M_F$. 
We thus find that Q-balls decay after the NLSP freezes out 
for $n=4$ with $\lambda \lesssim 10^{-8}$ 
and $n=6$.

Now, let us 
calculate the NLSP abundance from Q-ball decay, 
which is another source of gravitinos. 
A detailed analysis has been done in Ref.~\cite{Kasuya:2012mh} (see also Ref.~\cite{Kasuya:2011ix, Doddato:2012ja}),
where 
gravitino production from the direct decay of Q-balls 
via Planck-suppressed operators~\cite{ShKu, Kasuya:2011ix, Doddato:2011fz, Kasuya:2012mh} is taken into account. 
There, the ellipticity parameter $\epsilon$ is regarded as a free parameter. 
In this paper, we use $\epsilon = \mg/H_{\text{osc}}$ without fine-tuning.
In this case, 
gravitinos directly produced in Q-ball decay
is too inefficient to account for the observed DM density.

Q-balls decay into quarks at $T = \Td$ and lose their charges. 
Since the energy of Q-ball per unit charge, $\omega_0$, 
is proportional to $Q^{-1/4}$ (see Eq.~(\ref{omega_0 GMSB})), 
a Q-ball can decay into 
the next-to-lightest SUSY particle (NLSP)
once its charge decreases down to $Q_{\rm cr} \equiv (\sqrt{2} \pi M_F/\mN)^4$, 
where $\mN$ is the mass of the NLSP.%
\footnote{
In the case of $\bar{u} \bar{d} \bar{d}$ flat direction, 
Q-balls does not interact with sleptons and thus 
cannot decay into them. 
In this case, the mass of the NLSP for the condition $\omega_0 \lessgtr \mN$ in Eq.~(\ref{fr}) 
has to be replaced with that of the lightest MSSM particle which Q-balls can decay into. 
}
Thus we estimate the total number of the NLSP from the decay of each Q-ball as
\beq
 \text{fr}_{\text{N}} = 
 \left\{
 \bea{ll}
 \text{Br}_{\text{NLSP}} &\qquad \text{ for } \omega_0 > \mN, \\ 
 \text{Br}_{\text{NLSP}} \frac{Q_{\rm cr}}{Q} & \qquad \text{ for } \omega_0 < \mN,  
\eea
\right.
\label{fr}
\eeq
where the branching into the NLSP, $\text{Br}_{\text{NLSP}}$, is of the order of $0.01$. 
The first line is for the case that Q-balls can decay into the NLSP from the beginning, 
while the second one is for the case that Q-balls can decay into the NLSP only after 
their charges decrease down to $Q_{\rm cr}$. 
The combination of $Q_{\rm cr} / Q$ 
is rewritten as
\beq
 \frac{Q_{\rm cr}}{Q} &\equiv& 
 \frac{1}{Q} \lmk \frac{\sqrt{2}\pi M_F}{\mN} \rmk^4, \\
&\simeq&
   \left\{
  \bea{ll}
 3 
 \ k^{8/3}
 \lmk \frac{\mg}{2 \GEV} \rmk^{8/3}
 \lmk \frac{\lambda}{10^{-9}} \rmk^{4/3} 
 \lmk \frac{\mN}{5 \TEV} \rmk^{-4}, 
  &\qquad \text{ for } n = 4, \\
    5
 \ k^{16/5}
 \lmk \frac{\mg}{10 \GEV} \rmk^{16/5}
 \lmk \frac{\lambda}{10^{-4}} \rmk^{4/5} 
 \lmk \frac{\mN}{5 \TEV} \rmk^{-4}, 
  &\qquad \text{ for } n = 6. 
  \eea
  \right.
\eeq
We find that $Q_{\rm cr} / Q > 1$ for relatively large gravitino mass. 
The number density of the NLSP from Q-ball decay is then given as
\beq
 \frac{n_{\text{NLSP}}^{\text{decay}}}{s} \simeq
 \text{fr}_{\text{N}} \frac{n_\phi}{s}
 \simeq \text{fr}_{\text{N}} \frac{Y_b}{\epsilon b}. 
 \label{n_NLSP decay}
\eeq
If these NLSPs do not annihilate, 
the gravitino abundance from the Q-ball through the NLSP decay is given as 
\beq
 \frac{\rho_{3/2}^{\text{decay}}}{s} 
 &\simeq& \mg \text{fr}_{\text{N}} \frac{Y_b}{\epsilon b}, \\
 &\simeq& 
   \left\{
  \bea{ll}
 0.4 \EV
 \ k^{2/3}
 \lmk \frac{\text{fr}_{\text{N}} }{0.01} \rmk 
 \lmk \frac{\mg}{2 \GEV} \rmk^{2/3}
 \lmk \frac{\lambda}{10^{-9}} \rmk^{1/3}, 
  &\qquad \text{ for } n = 4, \\
    0.5 \EV
 k^{4/5}
 \lmk \frac{\text{fr}_{\text{N}} }{0.01} \rmk 
 \lmk \frac{\mg}{10 \GEV} \rmk^{4/5}
 \lmk \frac{\lambda}{10^{-4}} \rmk^{1/5}, 
  &\qquad \text{ for } n = 6. 
  \eea
  \right.
\label{rho DM from Q}
\eeq

The NLSPs produced from Q-balls 
are soon thermalized and may annihilate. 
We assume that the NLSP is right-handed stau
and hence its annihilation cross section is given by Eq.~(\ref{NLSP sigmav}). 
The annihilation effect of the NLSP results in the upper bound on the NLSP abundance
given by
\beq
  \frac{n_{\text{NLSP}}^{ \text{decay } (\text{ann}) }}{s}
 &\simeq&
 \sqrt{\frac{45}{8 \pi^2 g_*}} \frac{1}{\Td \la \sigma_{\text{ann}} v \ra_{\tilde{l}_R} \Mpl}, \\
 &\simeq& 
 \left\{
 \bea{ll}
 0.6 \EV
\ k^{-2/3}
\mg^{-1}
   \lmk \frac{\mg}{2 \GEV} \rmk^{1/3}
 \lmk \frac{\lambda}{10^{-9}} \rmk^{-5/6} 
 \lmk \frac{\mN}{5 \TEV} \rmk^2,  
 &\qquad \text{ for } n=4, \\
 0.4 \EV
\ k^{-1}
\mg^{-1}
 \lmk \frac{\lambda}{10^{-4}} \rmk^{-1/2} 
 \lmk \frac{\mN}{1 \TEV} \rmk^2,  
 &\qquad \text{ for } n=6. 
\eea
\right. 
 \label{n_NLSP ann}
\eeq
The gravitino abundance from the Q-ball through the NLSP decay 
is given by $\rho_{3/2}^{\text{decay (ann)}} = \mg n_{\text{NLSP}}^{\text{decay (ann)}}$.

\begin{figure}[t]
\centering 
\includegraphics[width=.45\textwidth, bb=0 0 236 235]{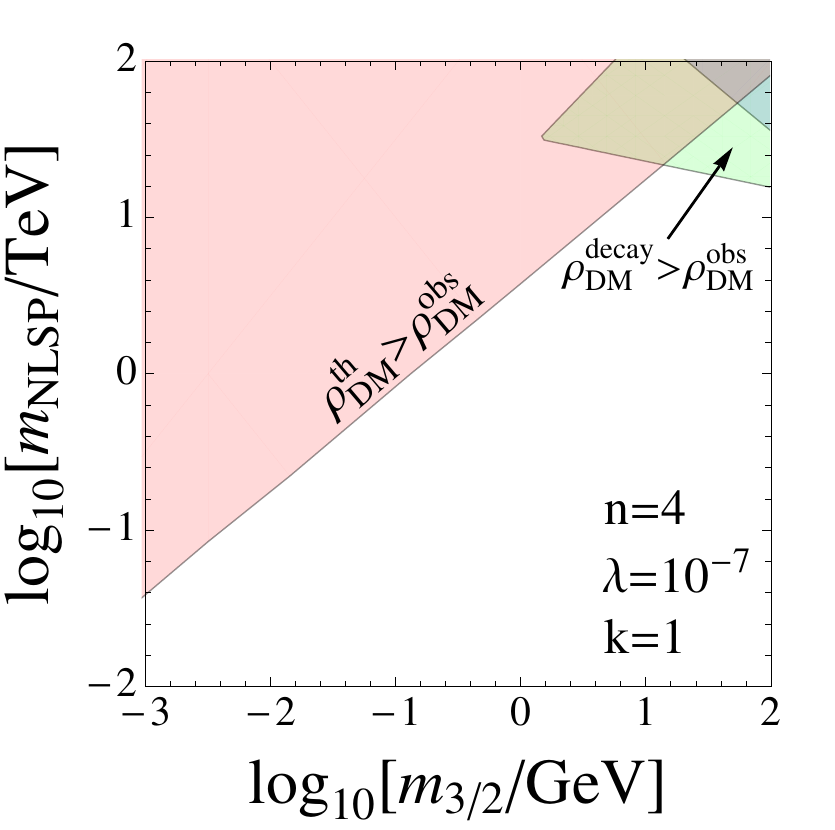} 
\hfill
\includegraphics[width=.45\textwidth, bb=0 0 240 239]{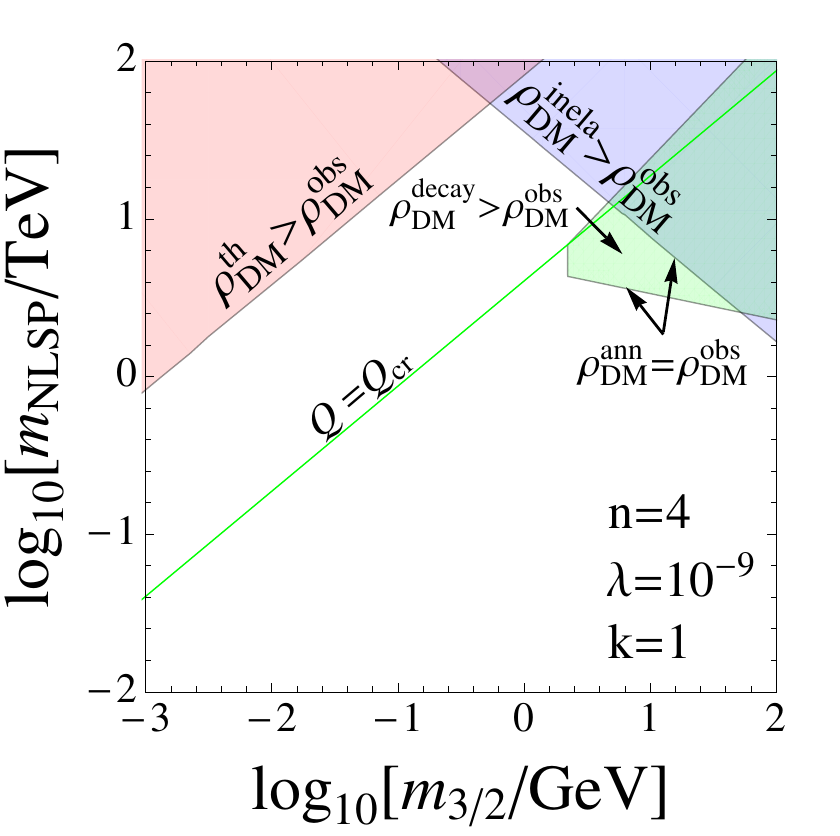} 
\caption{
Exclusion plot in a model of gauge mediation with Q-ball formation. 
We assume that 
$n=4$ and $k=1$. 
We also assume that 
$\lambda = 10^{-7}$ (left panel) and $10^{-9}$ (right panel). 
In the green (middle gray) shaded region, 
the abundance of DM
produced from 
Q-ball decay mediated by the NLSP ($\rho_{\rm 3/2}^{\rm decay}$) 
is larger than that observed. 
In the blue (dark gray) shaded region, 
the NLSP is produced too much through inelastic scatterings during the reheating epoch, 
that is, $\rho_{3/2}^{\rm inela} > \rho_{\rm DM}^{\rm obs}$. 
In the red (light gray) shaded region, 
the abundance of DM produced from the thermal plasma ($\rho_{\rm 3/2}^{\rm th}$) 
is larger than that observed. 
Here we have assumed that the mass of the gluino is five times larger than 
that of the NLSP. 
The lower boundaries of blue (dark gray) and green (middle gray) shaded regions 
are determined by the annihilation effect of the NLSP. 
The DM abundance is consistent with that observed 
at the boundary of the non-shaded regions. 
Q-balls can decay into the NLSPs from the first time (that is, $Q < Q_{\rm cr}$) 
below the green line. 
}
  \label{fig4}
\end{figure}

\begin{figure}[t]
\centering 
\includegraphics[width=.45\textwidth, bb=0 0 224 222]{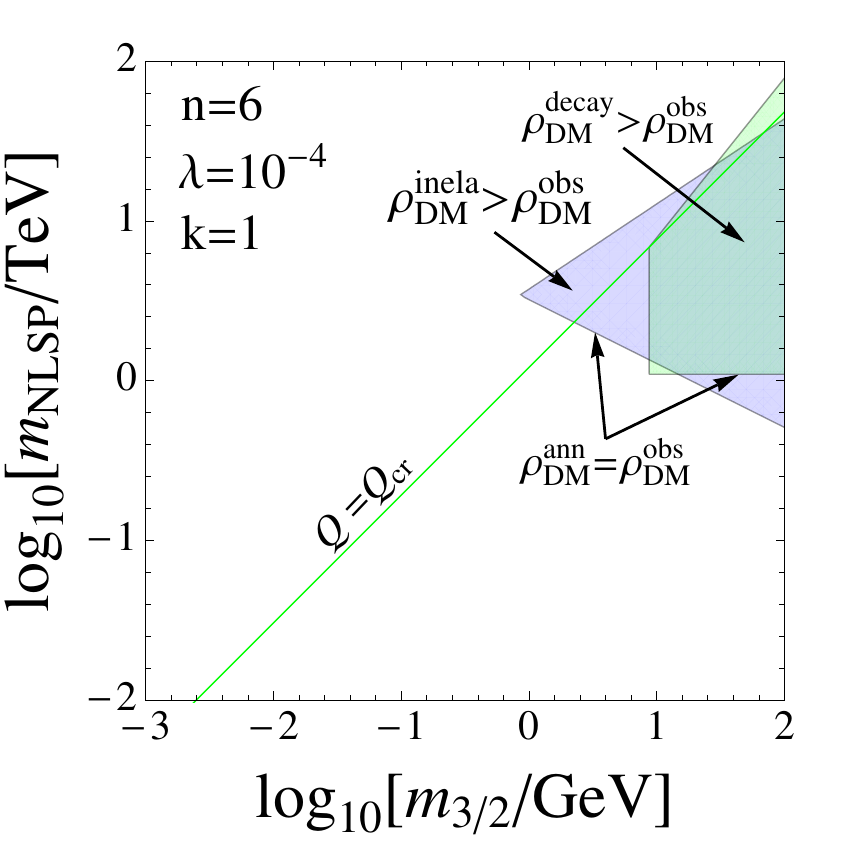} 
\hfill
\includegraphics[width=.45\textwidth, bb=0 0 231 230]{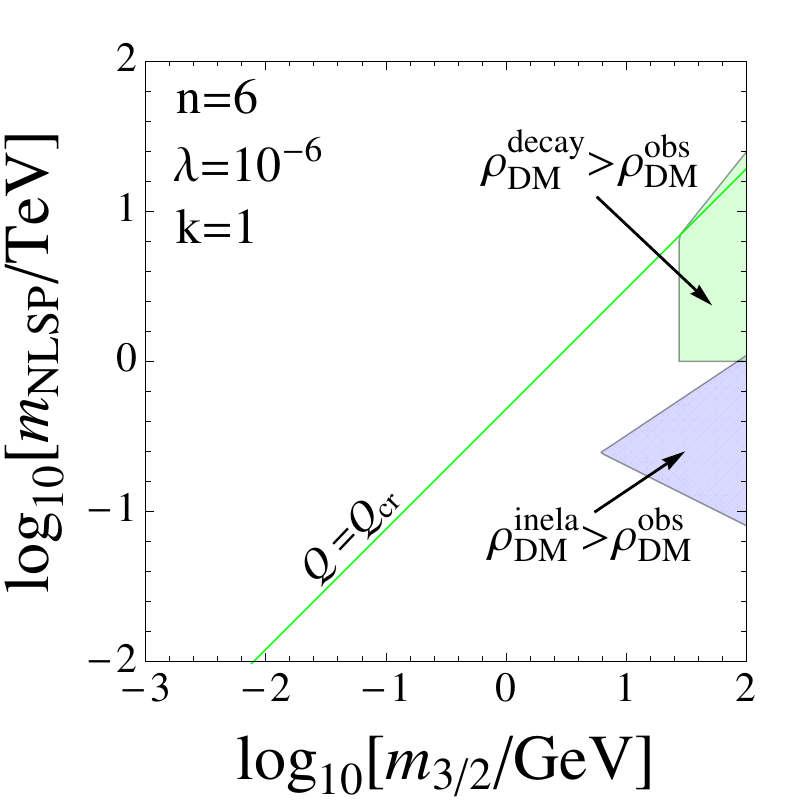} 
\caption{
Same as Fig.~\ref{fig4} but with $n=6$. 
We assume that 
$\lambda = 10^{-4}$ (left panel) and $10^{-6}$ (right panel). 
}
 \label{fig5}
\end{figure}

To sum up, 
the gravitino abundance is given by 
\beq
 \frac{\rho_{3/2}}{s} \simeq
 \frac{\rho^\text{th}_{3/2}}{s} + \min \lkk \frac{\rho^\text{inela}_{3/2}}{s}, 
 \frac{\rho^{\text{inela (ann)}}_{3/2}}{s} \rkk
 + 
 \text{min} \lkk \frac{\rho_{3/2}^{\text{decay}}}{s}, 
 \frac{\rho_{3/2}^{\text{decay } (\text{ann})}}{s} \rkk, 
\eeq
where $\rho^\text{th}_{3/2}$ is absent for the case of $n=6$. 
We show the results in Figs.~\ref{fig4} and \ref{fig5}. 
Inside blue (dark gray) shaded region, 
the gravitino abundance produced through inelastic scatterings during the reheating epoch 
is larger than the observed DM abundance. 
In the green (middle gray) shaded region, 
the abundance of DM
produced from 
Q-ball decay mediated by the NLSP ($\rho_{\rm 3/2}^{\rm decay}$) 
is larger than that observed. 
Q-balls can decay into the NLSPs from the first time (that is, $\omega_0 > m_{\text{NLSP}}$) 
below the green line. 
The lower boundary of 
the blue (dark gray) and green (middle gray) shaded regions 
are determined by the annihilation of the NLSP (Eqs.~(\ref{rho_anni_GM 2}) and (\ref{n_NLSP ann})). 
In the red (light gray) shaded region, 
the abundance of DM produced from the thermal plasma ($\rho_{\rm 3/2}^{\rm th}$) 
is larger than that observed. 
The DM abundance is consistent with that observed 
on the boundary of the non-shaded regions.

We find that 
the gravitino mass has to be larger than ${\cal O}(1) \GEV$ to account for the observed baryon 
and DM abundance for the case of $n=6$. 
For the case of $n=4$, 
while the DM abundance can be explained by the gravitino production from the thermal plasma 
in any value of gravitino mass, 
the gravitino production from Q-ball decay mediated by the NLSP 
can account for the DM abundance only if $\mg \gtrsim {\cal O}(1) \GEV$. 
As we have mentioned, this large value of the gravitino mass in general induces EDMs~\cite{Moroi:2011fi},
which will be detected in near future 
unless CP phases due to supergravity effect is suppressed by some reasons or tunings. 
In addition, 
since gravitinos are produced from the NLSP decay, 
the gravitino DM obtains a sizable free-streaming velocity derived by Eq.~(\ref{free streaming}). 
This again constraints the mass of the NLSP as 
$\mN \gtrsim 3 \TEV$~\cite{Viel:2013fqw}, 
and also 
this scenario would be tested by future observations of redshifted 21 cm line 
if $\mN \lesssim  14 \TEV$~\cite{Sitwell:2013fpa}.

\section{\label{conclusion}summary and conclusions}

We have proposed scenarios which account for the observed baryon asymmetry and DM density 
in models of gravity and gauge mediation, 
taking into account an implication on the energy scale of inflation $H_{\text{inf}} \simeq 10^{14} \GEV$ 
by the recent result of the BICEP2 experiment~\cite{Ade:2014xna}. 
We have considered the Affleck-Dine baryogenesis without Hubble induced $A$-term potentials, 
which is 
indeed the case 
for most models of high-scale inflation in supergravity~\cite{Kasuya:2008xp, 
Kawasaki:2000yn, Kallosh:2010ug} 
and for $D$-term inflation~\cite{Enqvist:1998pf, Enqvist:1999hv, Kawasaki:2001in}. 
In this case, the Affleck-Dine baryogenesis is excluded by 
an experimental upper bound on baryonic isocurvature perturbation, 
unless the initial phase of the AD field is tuned or 
the VEV of the AD field is as large as the Planck scale during inflation. 
We have investigated the latter possibility in detail.

Since the AD field with large VEV
results in larger energy density ratio for AD field/inflaton, 
it indicates
lower reheating temperature of the Universe to account for the present
baryon density successfully without additional entropy production. 
Hence, this scenario predicts a relatively low reheating temperature.
When the reheating temperature is low, DM is dominantly produced in non-thermal processes.
Therefore, it is necessary to investigate the non-thermal processes in detail.
In addition, the Affleck-Dine baryogenesis often results in 
formation of 
Q-balls, 
which decay into light SUSY particles as well as quarks at later time. 
Based on these issues, 
we have constructed consistent scenarios to account for the observed baryon and DM densities 
in the cases with and without Q-ball formation in models of gravity and gauge mediation.

In gravity mediation, DM is produced mainly from two sources: 
direct decay of inflaton into MSSM particles~\cite{Moroi:1994rs, Kawasaki:1995cy, 
Moroi:1999zb, Allahverdi:2002nb, Gelmini:2006pw, Kurata:2012nf}
and inelastic scatterings during reheating process~\cite{Allahverdi:2002nb, HKMY}. 
While the former process depends on unknown inflaton mass, 
the latter process depends only on the reheating temperature and DM mass. 
From the observation of the DM abundance, 
we have predicted the DM mass around the TeV scale. 
In addition, if Q-balls are formed after the Affleck-Dine baryogenesis, 
they emit gauginos lighter than squarks and thus they can be another source of DM. 
Interestingly, since the Pauli blocking effect at the Q-ball surface gives 
a simple relation between the branching into quarks and gauginos from Q-ball decay, 
we can overcome the baryon-DM coincidence problem in this scenario~\cite{KKY}. 
We have predicted the mass of the bino, which is the LSP in typical gravity mediation models, 
as ${\cal O}(1) \TEV$. 
These scenarios would be 
tested by direct detection experiments in the near future~\cite{Allahverdi:2002nb}. 

In gauge mediation, gravitino DM is produced by scatterings between 
gluon and gluino in the thermal bath at the reheating epoch~\cite{Moroi:1993mb}. 
We have confirmed that the baryon and DM density can be simultaneously explained 
in the Affleck-Dine baryogenesis when the mass of the gravitino is about ${\cal O}(1) \MEV$ or ${\cal O}(1) \GEV$, 
depending on the power of superpotential for the AD field. 
However, any flat directions other than $L H_u$ 
fragment into Q-balls
after the Affleck-Dine baryogenesis. 
Then 
Q-balls decay into NLSPs as well as quarks and the NLSPs eventually decay into 
gravitinos~\cite{Doddato:2012ja, Kasuya:2012mh}. 
We have found that this scenario can also explain the observed DM density,
and have predicted that the gravitino and the NLSP masses are larger than about ${\cal O}(1) \GEV$ 
and 
$3 \TEV$, 
respectively.
The lower bound on the NLSP mass comes from an upper bound on the present-day
free-streaming velocity of DM. 
This scenario would be tested by future observations of redshifted 21 cm line 
if the NLSP mass is less than about $14 \TEV$.
The gravitino mass of ${\cal O}(1) \GEV$ induces detectable EDMs in near future experiments unless CP phases due to supergravity effect is suppressed by some reasons or tunings.

%
\section*{Acknowledgements}
K.H. thanks Norimi Yokozaki for useful discussion. 
This work is supported by
a Grant-in-Aid for Scientific Research 
from the Ministry of Education, Science, Sports, and Culture (MEXT), Japan, No. 25400248 (M.K.) and No. 21111006 (M.K.),
the World Premier International Research Center Initiative (WPI Initiative), 
MEXT, Japan (A.K., M.K. and M.Y.),
the Program for Leading Graduate Schools, MEXT, Japan (M.Y.),
and JSPS Research Fellowships for Young Scientists (K.H., K.M. and M.Y.).
%

\appendix*
\section{\label{appendix}Q-ball decay rate into massive particles}

In this appendix, we discuss the Q-ball decay rate into a massive particle $\chi$~\cite{KKY}. 
While flux of a massless particle at the Q-ball surface is calculated as in Eq.~(\ref{flux}), 
that of a massive particle is suppressed by its mass.
Here we consider the Q-ball decay through an elementary process 
$\tilde{q} \to q + \chi$.
We denote the mass of $\chi$ as $m_{\chi}$. 
Since the total energy of this process is given by the energy of the Q-ball per unit charge, $\omega_0$, 
the particle $\chi$ obtains energy in the range of $[m_{\chi}, \omega_0]$
and the quark obtains energy in the range of $[0, \omega_0-m_{\chi}]$. 
Their flux is determined by the following procedure. 
Due to conservation of energy and angular momentum, 
quark flux with the energy of $E$
have to coincide with 
$\chi$ flux with the energy of $\omega_0 - E$. 
Since either of them cannot exceed upper bound on their flux due to the Pauli blocking effect, 
their flux is determined by the severer bound. 
The quark flux with the energy of $E$ is proportional to $\dd p_{\rm quark} = \dd E$, 
while the $\chi$ flux with the energy of $\omega_0 - E$ 
is proportional to $v_\chi \times  \dd p_{\chi} = p_{\chi} / E \times E/p_{\chi} \dd E = \dd E$, 
where we use $p_{\chi}^2 = (\omega_0 - E)^2 - m_\chi^2$. 
We obtain their flux at the Q-ball surface as 
\beq
 \bm{n \cdot j_\chi} 
 \simeq
  \frac{1}{8 \pi^2}
  \int^{\omega_0 - m_{\chi}}_0 \text{d}E
    \min \lkk E^2, (\omega_0 - E)^2 - m_\chi^2 \rkk. 
\eeq
This integral can be performed analytically
and we obtain 
the following
correction to the flux given in Eq.~(\ref{flux}): 
\begin{eqnarray}
  \bm{n\cdot j}_{\chi} & \simeq & \frac{\omega_0^3}{96 \pi^2} 
  \times f(m_{\chi}/\omega_0), \label{bino massive}\\[0.6em]
f(x ) &\equiv& 
	\begin{cases}
	  1- 6 x^2 + 8 x^3 - 3 x^4 &\text{for}~~ 0 \le  x \le 1 \\
	  0  &\text{for}~~ 1 <  x
	 \end{cases}
	  \label{correction}
\end{eqnarray}
Note that while 
the flux of sparticles is given by the above formula,
the flux of quarks takes a different form 
$8 \times \omega_0^3 / (96 \pi^2)$, 
because they are produced through scattering process inside the Q-ball
(see the discussion below Eq.~(\ref{flux}).



\end{document}